\documentclass[]{aa}

\usepackage{graphicx}
\usepackage{txfonts}
\usepackage{hyperref}
\usepackage{xcolor}
\hypersetup{
colorlinks,
linkcolor={red!80!black},
citecolor={blue!80!black},
urlcolor={blue!80!black}
}

\newenvironment{myfont}{\fontfamily{ppl}\selectfont}{\par}
\DeclareTextFontCommand{\textmyfont}{\myfont}

\newcommand{\code}[1]{\texttt{#1}}

\def\nifs{\iso{56}Ni}
\def\cofs{\iso{56}Co}

\def\cm3{cm$^{-3}$}
\def\kms{\mbox{km~s$^{-1}$}}
\def\msunyr{$M_{\odot}$\,yr$^{-1}$}

\def\msun{$M_{\odot}$}

\def\one{\ts {\,\sc i}}
\def\two{\ts {\,\sc ii}}

\def\beq{\begin{equation}}
\def\eeq{\end{equation}}

\def\lesssim{\mathrel{\hbox{\rlap{\hbox{\lower4pt\hbox{$\sim$}}}\hbox{$<$}}}}
\def\gtrsim{\mathrel{\hbox{\rlap{\hbox{\lower4pt\hbox{$\sim$}}}\hbox{$>$}}}}

\def\one{{\,\sc i}}
\def\two{{\,\sc ii}}

\def\v1d{{\code{V1D}}}

\def\obs2d{{\code{OBS\_2D}}}
\def\cmfgen{{\code{CMFGEN}}}

\def\ergs{erg\,s$^{-1}$}

\def\oiidoub{[O\two]\,$\lambda\lambda$\,$7319,\,7320$}
\def\mgi{Mg\one]\,$\lambda\,4571$}
\def\oidoub{[O\one]\,$\lambda\lambda$\,$6300,\,6364$}
\def\caiidoub{[Ca\two]\,$\lambda\lambda$\,$7291,\,7323$}

\def\niidoub{[N\two]\,$\lambda\lambda$\,$6548,\,6583$}
\def\niiauroral{[N\two]\,$\lambda$\,$5755$}
\def\mgiires{Mg\two\,$\lambda\lambda$\,$2796,\,2803$}

\newcommand{\iso}[2]{\ensuremath{^{#1}\rm{#2}}}

\begin{document}

\title{The morphing of decay powered to interaction powered Type II supernova ejecta at nebular times
}

\titlerunning{The morphing of decay powered to interaction powered Type II SNe}

\author{
Luc Dessart\inst{\ref{inst1}}
\and
Claudia P. Guti\'errez\inst{\ref{inst2},\ref{inst3}}
\and
Hanindyo Kuncarayakti\inst{\ref{inst3},\ref{inst2}}
\and
Ori D. Fox\inst{\ref{inst4}}
\and
Alexei V. Filippenko\inst{\ref{inst5}}
  }

\institute{
     Institut d'Astrophysique de Paris, CNRS-Sorbonne Universit\'e, 98 bis boulevard Arago, F-75014 Paris, France\label{inst1}
 \and
     Finnish Centre for Astronomy with ESO (FINCA), FI-20014 University of Turku, Finland\label{inst2}
\and
     Tuorla Observatory, Department of Physics and Astronomy, FI-20014 University of Turku, Finland\label{inst3}
 \and
     Space Telescope Science Institute, 3700 San Martin Drive, Baltimore, MD 21218, USA\label{inst4}
 \and
     Department of Astronomy, University of California, Berkeley, CA, 94720-3411, USA\label{inst5}
   }

   \date{}

  \abstract{
Much excitement surrounds the intense mass loss that seems to take place in some massive stars immediately before core collapse. However, occurring too late, it has a negligible impact on the star's evolution or the final yields, which are influenced instead by the longer-term, quasi-steady mass loss taking place during H and He burning. Late-time observations of core-collapse supernovae interacting with the progenitor wind are one means to constrain this secular mass loss. Here, we present radiative transfer calculations for a Type II SN with and without this interaction power, focusing on the phase between 350 and 1000\,d after explosion. Without interaction power, the ejecta are powered through radioactive decay whose exponential decline produces an ever-fading SN. Instead, with a constant interaction power of $10^{40}$\,\ergs\ (representative of an SN~II ramming into a steady-state $10^{-6}$\,\msunyr\ wind), the spectrum morphs from decay powered at 350\,d, with narrow lines forming in the inner metal-rich ejecta, to interaction powered at 1000\,d, with broad boxy lines forming in the outer H-rich ejecta. Intermediate times are characterized by a hybrid and complex spectrum made of overlapping narrow and broad lines. While interaction boosts primarily the flux in the ultraviolet, which remains largely unobserved today, a knee in the $R$-band light curve or a $U$-band boost are clear signatures of interaction at late times. The model predictions compare favorably with a number of Type II supernovae including SN\,2004et or SN\,2017eaw at 500--1000\,d after explosion.
  }

   \maketitle
%

\section{Introduction}

Mass loss is a ubiquitous phenomenon affecting all massive stars. While generally weak during the main sequence, stellar wind mass loss rises as the star evolves and becomes more luminous. In a single star of 15\,\msun\ initially, it leads to a cumulative mass reduction of $\sim$\,10\,\%, but this fraction may rise up to 90\% for the highest mass stars by the time they reach core collapse \citep{maeder_meynet_87,langer_massive_94}. Mass transfer in interacting binaries leads to short-lived but stronger mass loss episodes, often causing the loss of the H-rich envelope and leading to H-deficient progenitors at core collapse \citep{podsiadlowski_92,wellstein_langer_99,eldridge_08_bin,langer_araa_mdot_12}. In such binaries, stellar wind mass loss also operates and may influence the frequency of Type Ib versus Type Ic SNe \citep{yoon_wr_17,aguilera_dena_ibc_22} (see \citealt{filippenko_rev_97} for a review of SN classification). Instead of constraining mass loss through the observation of the atmospheres and environments of stars at various stages of their lives, one may monitor core-collapse supernovae (SNe) for years and decades after explosion when the ejecta expand into and interact with the material lost by the star during the last hundreds or thousands of years before collapse.  For a SN shock ramming at velocity $V_{\rm sh}$ into a steady-state wind with mass-loss rate $\dot{M}$ and velocity $V_\infty$, the instantaneous power released by the interaction is
$L_{\rm sh} = \dot{M} V_{\rm sh}^3 / 2 V_\infty
= 3.15 \times 10^{40} \,\,\, \dot{M}_{-5} V_{\rm sh,4}^3 / V_{\infty,2} \,\,\, {\rm erg}\,{\rm s}^{-1} \, ,
$
where $\dot{M}_{-5} \equiv \dot{M} / 10^{-5} M_\odot$ yr$^{-1}$, $V_{\rm sh,4} \equiv V_{\rm sh}$/10,000\,km\,s$^{-1}$, and $V_{\infty,2} \equiv V_\infty/100$\,km\,s$^{-1}$. Hence, even modest wind strengths rival with radioactive decay power at late times \citep{DH_interaction_22}. Unfortunately, such late-time observations of SNe are rare, and most SN observations even today remain focused on the earlier, brighter phase of the SN evolution.

The observational signatures suggestive of ejecta interaction with circumstellar material (CSM) are diverse. The interaction may produce radio and X-ray emission, as  observed over a year-time scale in the luminous fast-declining Type II SN\,1979C (\citealt{chevalier_82, chevalier_82_radio_xray}; \citealt{weiler_79c_91}). In some objects, this emission is observed with a delay, as in SN\,2014C \citep{margutti_14C_16}, suggesting interaction with material not directly at the progenitor surface but far away from it, at $10^{16}-10^{17}$\,cm. Strong ultraviolet (UV) emission is also identified, in particular in strongly interacting events like the Type IIn SN\,2010jl  \citep{fransson_10jl} or the Type Ibn SN\,2006jc \citep{immler_06jc_08}, but UV emission is typically expected in all ejecta/wind interactions \citep{fransson_uv_84}. Because of an observational bias, spectroscopic signatures of interaction are mostly confined to the optical range. Early-time emission profiles with narrow cores and symmetric broad wings are the defining signature of Type IIn SNe. This profile morphology may persist for many days (e.g., SN\,1998S, \citealt{leonard_98S_00}, or SN\,2020tlf, \citealt{wynn_20tlf_22}) or only for hours (e.g., SN\,2013fs; \citealt{yaron_13fs_17}), corresponding to different extent and mass of the CSM. In some Type II SNe, a high-velocity absorption feature is observed in some lines, in particular in H$\alpha$, which suggests the presence of a dense shell at the interface between ejecta and CSM \citep{chugai_hv_07,gutierrez_pap1_17,DH_interaction_22}.  Ejecta/CSM interaction may also boost the  emissivity of the outer ejecta. One such manifestation is observed in the optical spectra of the Type IIb SN\,1993J, which revealed broad boxy emission profiles throughout the nebular phase and detectable for years after explosion \citep{matheson_93j_00a,matheson_93j_00b}. Since SN\,1993J, other SNe have shown broad, sometimes boxy emission, although generally limited to H$\alpha$, and with a turn-on occurring at different times: at $\sim$\,200\,d in SNe 2007od, 2017ivv, and 2014G \citep{andrews_07od_10,gutierrez_17ivv_20,terreran_14G_16}, at $\sim$\,450\,d in SN\,2013by \citep{black_13by_17}, or at $\sim$\,600\,d in SNe 2007it and 2017eaw \citep{andrews_07it_11,weil_17eaw_20}.

On the theoretical and modeling side, most work has been focused on early-time signatures of interaction and superluminous SNe -- that is, interaction with massive CSM corresponding to very large mass-loss rates of order 10$^{-1}$\,\msunyr\ as inferred for SN\,2010jl. But theory predicts that massive stars may undergo episodes of intense mass loss, especially in interacting binaries, producing intermediate states between mass-loss rates of 10$^{-6}$\,\msunyr\ typical of the red-supergiant (RSG) star Betelgeuse today \citep{dolan_betelgeuse_16} and the most extreme events like SN\,2010jl \citep{fransson_10jl,D15_2n}. In most cases, the pre-SN mass-loss rates would be too small to produce an optically thick CSM, hence the transient would never show (or at best very briefly after shock breakout) the narrow spectral signatures that flag the event as Type IIn \citep{DH_interaction_22}. Yet, the power from interaction could be substantial, rival with and even supersede the contribution from decay power at late times (see, for example, indications for this based on the H$\alpha$ line strength in \citealt{chugai_late_sn2p_90}). Signatures of interaction should be present at some level and with much diversity in essentially all core-collapse SNe -- it should be the norm rather than the exception.

The impact of ejecta/CSM interaction in a standard RSG star explosion model was recently explored by \citet{DH_interaction_22}. To mimic a range in steady-state pre-SN wind mass-loss rate from a value vanishingly small up to about $10^{-3}$\,\msunyr, shock power was introduced in the outer ejecta in a dense shell of 0.1\,\msun\ located at $\sim$\,11,000\,\kms\ and with a constant rate between zero and 10$^{43}$\ergs. The simulations were evolved from 15\,d until 350\,d, thus covering the photospheric and the nebular phase. In the model without shock power, a high-velocity absorption feature developed at the recombination epoch in numerous lines including H$\alpha$, H$\beta$, Na\one\,D, O\one\,7774\,\AA, or the Ca\two\ near-infrared (IR) triplet. However, with shock power, the models exhibited a weak boost in continuum optical flux, a broad boxy H$\alpha$ emission component, and a boost to the UV flux which strengthened with time, the more so for greater shock power. All these properties are compatible with the observations, but the models suggest that the current sample of observations is capturing only weakly the full breadth of interaction signatures and these signatures should also be more frequently detected. The spectacular evolution captured for SN\,1993J over a decade after discovery was made possible in part because of its proximity, but the pre-SN mass loss was inferred to be modest, with a value of only a few times 10$^{-5}$\,\msunyr\ (e.g., \citealt{fransson_bjornsson_93J_98}; \citealt{tatischeff_93J_09}). Current 8-m-class telescopes and future facilities such as the Vera Rubin Observatory should be able to capture signatures of interaction in all core-collapse SNe (though depending on the SN distance) and help constrain the pre-SN mass-loss rate of massive stars, whether it is compatible with single-star or binary-star evolution, and so on.

In this paper, we extend the study of \citet{DH_interaction_22}, which covered epochs prior to 350\,d, and focus on the evolution from 350 to 1000\,d of a Type II SN model with and without shock power. The qualitative evolution is similar for a wide range of interaction powers, so we focus on only one case with an injected shock power of 10$^{40}$\ergs, which should be representative of a standard Type II SN progenitor and explosion. In other words, any RSG star explosion should be subject to that power level and thus our results are representative, default expectations for a standard SN II. Weaker values would suggest even lower mass-loss rates, which would imply a negligible impact of wind mass loss on the progenitor evolution. In the next section, we start by emphasizing how the exponential decline of radioactive decay should eventually be swamped by ejecta/CSM interaction at late times. In Section~\ref{sect_setup}, we describe the models and setup for the present calculations. Section~\ref{sect_res} discusses the results for the photometric and spectroscopic evolution for the models with and without shock power, covering from 350\, until 1000\,d. In Section~\ref{sect_obs}, we compare these results to a few well-observed events exhibiting signatures of interaction. The implications as well as the limitations of our results are discussed in Section~\ref{sect_disc}. We present our conclusions in Section~\ref{sect_conc}.

\begin{table}
    \caption{Summary of ejecta properties for model s15p2 \citep{sukhbold_ccsn_16,D21_sn2p_neb}. Numbers in parenthesis represent powers of ten.
\label{tab_prog}
}
\begin{center}
\begin{tabular}{lcc}
\hline
Quantity   &   Unit          & Value \\
\hline
$M_{\rm init}$  & [\msun] & 15.2      \\
$M_{\rm ej}$ & [\msun] &  10.95   \\
$E_{\rm kin}$ & [erg]  &     8.37(50)   \\
  H-env & [\msun] & 8.13    \\
  \hline
  \multicolumn{3}{c}{Stable isotopes} \\
  \hline
   \iso{1}H & [\msun] &5.24   \\
  \iso{4}He & [\msun] &3.93   \\
  \iso{12}C & [\msun] &1.7(-1)    \\
  \iso{16}O & [\msun] & 1.0    \\
  \iso{24}Mg  & [\msun] & 4.41(-2)    \\
  \iso{28}Si &  [\msun] &8.65(-2)    \\
  \iso{40}Ca & [\msun] & 7.77(-3)   \\
\hline
  \multicolumn{3}{c}{Unstable isotopes} \\
  \hline
 \iso{44}Ti & [\msun] & 4.31(-5)     \\
 \iso{48}Cr & [\msun]& 1.53(-4)   \\
 \iso{49}Cr&  [\msun]& 4.19(-6)    \\
\iso{51}Mn & [\msun]& 1.12(-5)    \\
\iso{52}Fe & [\msun] & 6.95(-4)     \\
\iso{55}Co&  [\msun] & 3.20(-4)   \\
\iso{56}Ni & [\msun] & 6.33(-2)    \\
\iso{57}Ni&  [\msun] & 2.79(-3)    \\
\hline
\end{tabular}
\end{center}
\end{table}

\begin{figure}
\centering
\includegraphics[width=\hsize]{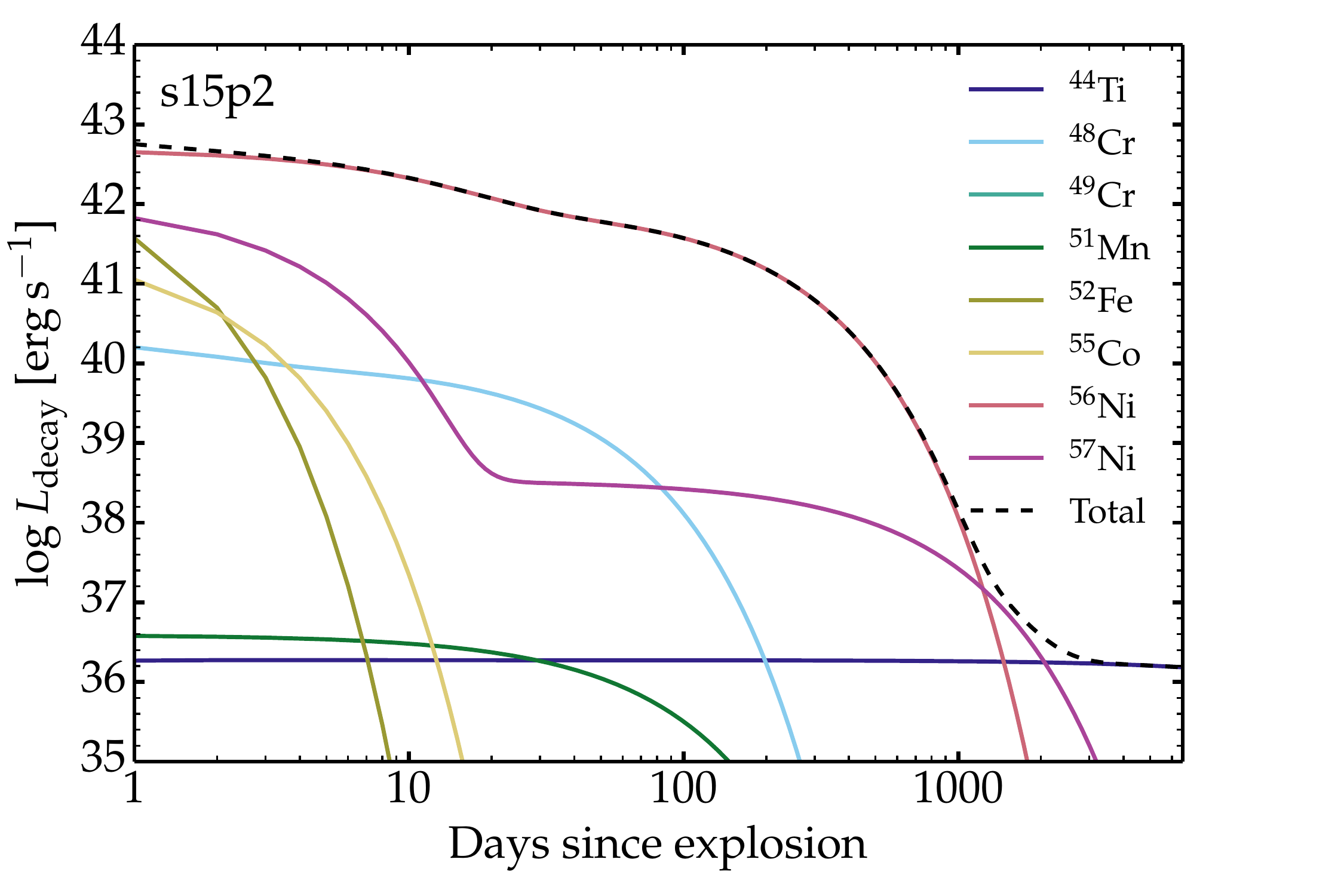}
\vspace{-0.4cm}
\caption{Emitted decay power as a function of time since explosion in model s15p2 (see Table~\ref{tab_prog} for the yields). We show the individual contributions from the dominant unstable isotopes (the label refers to the parent isotope of the two-step decay chains) as colored curves, with the total given by the black dashed curve.   Because of $\gamma$-ray escape, the power effectively absorbed by the ejecta is only a fraction of the power emitted (about 25\% at 1000\,d in this model). The dominant source of power is \nifs\ in the first 1000\,d, while $^{44}$Ti takes over beyond about 3000\,d. Evidently, ejecta interaction with a standard RSG wind, which delivers a power of about 10$^{40}$\,\ergs\ would dominate over radioactive decay after about 600\,d.
\label{fig_ldecay}
}
\end{figure}

\section{All core-collapse SNe should eventually be powered by interaction}

Figure~\ref{fig_ldecay} illustrates the evolution of the decay power emitted by the most abundant unstable isotopes produced in the explosion of a star of 15.2\,\msun\ initially \citep{sukhbold_ccsn_16}, namely and in order of increasing atomic mass \iso{44}Ti, \iso{48}Cr, \iso{49}Cr, \iso{51}Mn, \iso{52}Fe, \iso{55}Co, \iso{56}Ni, and \iso{57}Ni. The associated yields are listed in Table~\ref{tab_prog} together with other ejecta properties. These parent isotopes are part of a two-step decay chain. In this model, which is representative of a standard core-collapse SN in terms of explosive yields, \nifs\ is the main source of decay power up until  $\sim$\,1000\,d, followed by an intermediate phase during which the decay of \iso{57}Ni dominates, until the decay of \iso{44}Ti takes over after about 3000\,d. During the first 100\,d after explosion, the power released from \nifs\ is of order 10$^{42}$\,\ergs\ and thus comparable to a Type II SN luminosity. When \iso{57}Ni takes over, the total decay power emitted is only 10$^{37}$\,\ergs, after which the slow-decaying \iso{44}Ti isotope (which has a half-life of 21,915\,d) delivers a near-constant power of about 10$^{36}$\,\ergs\ (a maximum time of 7000\,d is used in Fig.~\ref{fig_ldecay}).

Such powers can easily be matched at late times by ejecta interaction with CSM. A standard secular steady-state mass-loss rate of 10$^{-6}$\,\msunyr\ in an RSG should deliver a power of about 10$^{40}$\,\ergs, hence greater than the total decay power in  Fig.~\ref{fig_ldecay} at all times $>$\,600\,d. The major uncertainty here is how much of that power thermalizes in the outer ejecta to escape as lower energy photons, and what fraction of these come out in the UV and in the optical (a further aspect is whether the radiation emerges in the continuum or in lines; \citealt{DH_interaction_22}). This will depend on the previous history of the ejecta (i.e., the presence or absence of a dense shell in the outer ejecta, which will form through the continuous pile-up of wind material \citep{chugai_hv_07}, but also directly at shock breakout \citep{d17_13fs}) and the magnitude of the RSG wind mass-loss rate. If the shock power is not thermalized, the SN should be X-ray luminous. Somehow, this power has to come out, and it is not small, even for a modest wind mass loss.

However, SN\,1987A is one example that demonstrates that it may take a very long time before interaction with CSM becomes the dominant power source. Indeed, modeling of the nebular-phase spectrum of SN\,1987A suggests that its ejecta were powered primarily by the decay of \iso{44}Ti at 8\,yr after explosion \citep{jerkstrand_87a_11}. The low late-time power inferred from optical and near-IR radiation suggests that the CSM density is low, which may in part be due to the large wind velocity of the blue-supergiant progenitor relative to the RSG case.

\section{Numerical setup}
\label{sect_setup}

  The simulations with the time-dependent nonlocal thermodynamic equilibrium radiative transfer code 
 \cmfgen\ \citep{HD12} presented by \citet{DH_interaction_22} were focused on the earlier-time evolution of Type II SN ejecta under the influence of shock power. In that case, the issue of chemical mixing is not critical and thus a simple boxcar mixing, which introduces both macroscopic and microscopic mixing, was used. Here, we focus on the late time evolution when the treatment of mixing is critical. We thus use the shuffling method of \citet{DH20_shuffle} and employ the detailed 15.2\,\msun\ model of \citet{sukhbold_ccsn_16}, with the numerical setup presented by \citet[the model is named s15p2]{D21_sn2p_neb}. To avoid repetition, and since the focus in this paper is on the influence of interaction power on the SN radiation rather than the impact of chemical mixing, we refer the reader to \citet[in particular Section~2]{D21_sn2p_neb} for additional details.

  Here, we evolve this model s15p2 from the initial time of 350\,d until 1000\,d after explosion accounting for the radioactive decay from \nifs\ and \cofs. A first model, named s15p2NoPwr, neglects any contribution from shock power. A second model, named s15p2Pwr1e40, is evolved with an extra power of 10$^{40}$\,\ergs\ injected at 8000\,\kms\ within a dense shell of 0.1\,\msun. This power corresponds to the thermalized  part of the shock power arising from interaction with CSM, which for this value could arise from a pre-SN wind of about 10$^{-6}$\,\msunyr\ in the case of efficient thermalization. To mimic the potential breakup of the shell in three dimensions, we spread the dense shell over a thickness of about 10\% of the shell radius and introduce clumping to maintain a high gas density within the shell. The clumping profile follows a Gaussian with a minimum volume filling factor of 1\% at the shell center. Ultimately, the exact structure of this dense shell requires three-dimensional (3D) radiation-hydrodynamics simulations of the ejecta/CSM interaction, and such simulations are currently unavailable.

We performed a few additional simulations for model s15p2 at 350\,d by varying the injected shock power from $1 \times 10^{40}$ to $7 \times 10^{40}$\,\ergs. These simulations are analogous to model s15p2Pwr1e40 at 350\,d. They are used for comparison to observations in Section~\ref{sect_obs} and shown in the Appendix in Fig.~\ref{fig_other_pwr}.

In the models with shock power, 70 extra grid points were needed to resolve the dense, clumped shell. Models with (without) shock power used 420 (350) grid points and each model in the sequence took about a week to reach convergence (this is aggravated by the need for a small turbulent velocity of 10\,\kms\ in the \cmfgen\ calculation; \citealt{DH20_neb}). Because of this cost, it is not practical to run many simulations. At earlier times, the simpler mixing can be used and only 100 grid points are needed \citep{DH_interaction_22}. At times later than 1000\,d (or earlier in Type IIb, Ib, or Ic SN ejecta) one may focus only on the outer dense shell and thus use only about 70 points. At intermediate nebular epochs of 1--3\,yr, which are the focus of the present study, an accurate calculation requires high resolution for both the inner ejecta mixed with the shuffling technique and the narrow outer dense shell where the shock power is deposited.

\section{Results}
\label{sect_res}

\subsection{Ejecta properties}

\begin{figure}[t]
\centering
\includegraphics[width=\hsize]{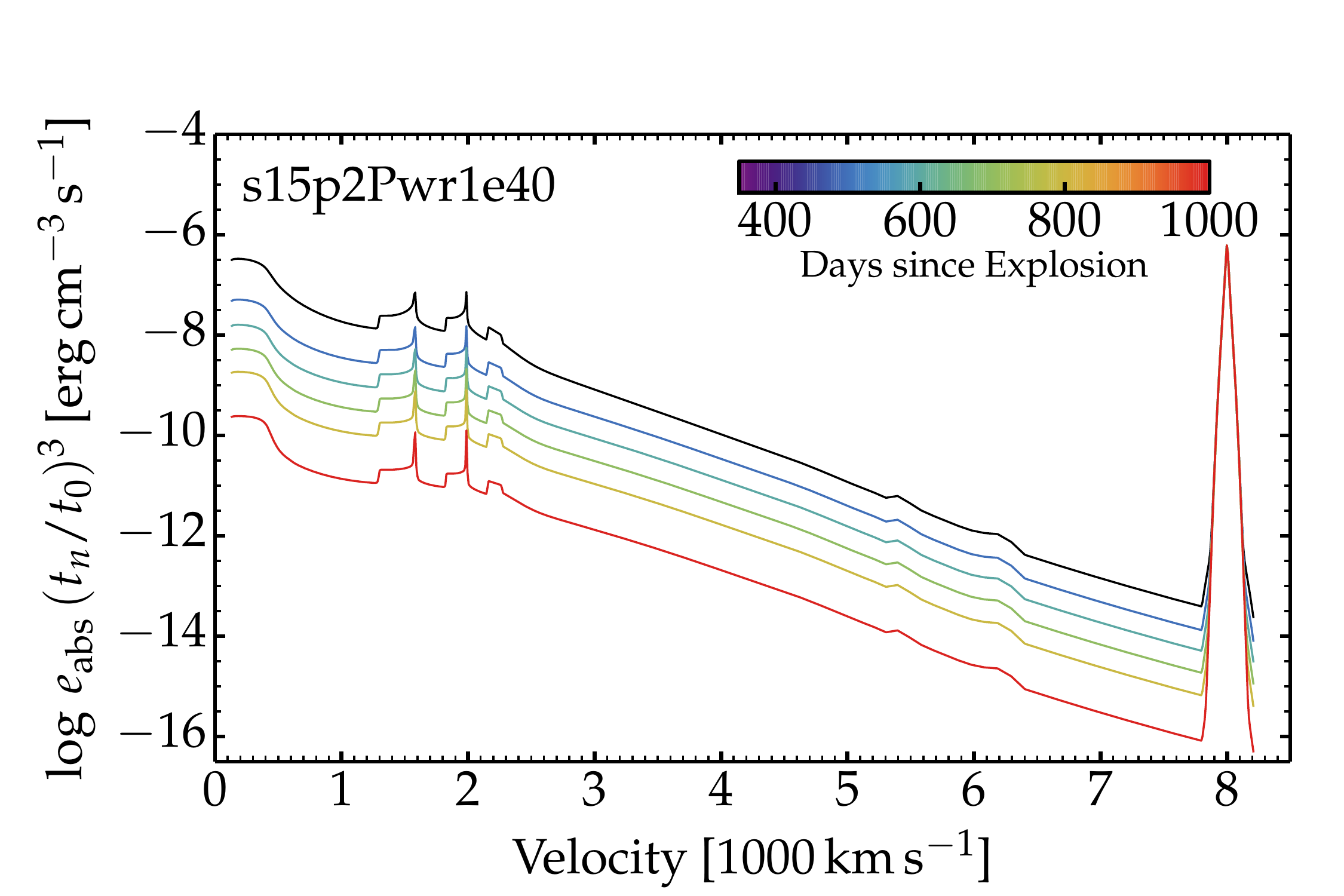}
\includegraphics[width=\hsize]{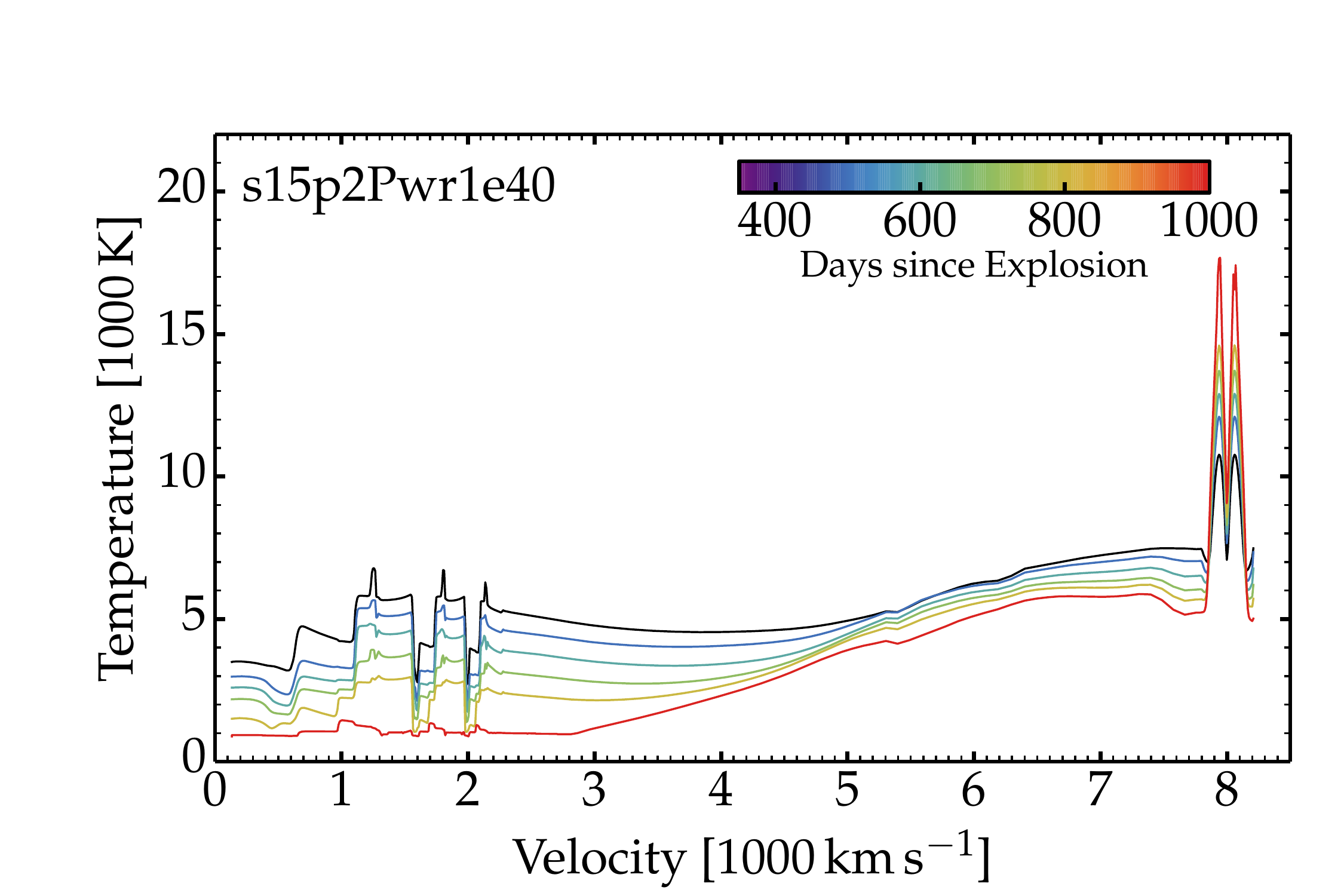}
\includegraphics[width=\hsize]{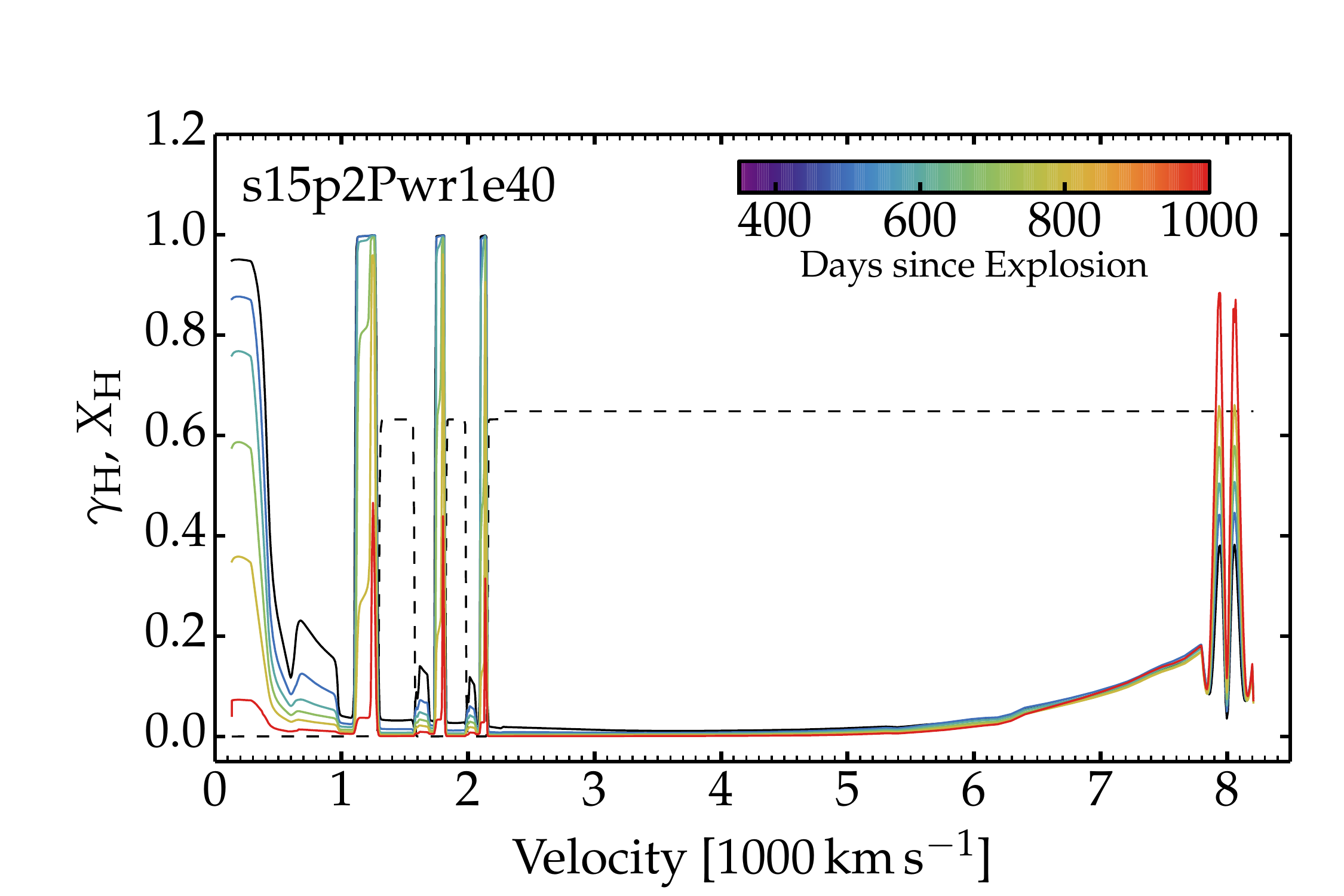}
\caption{Evolution of ejecta properties for model s15p2Pwr1e40 from $t_0=$\,350 to 1000\,d. From top to bottom, we show the profile of the total power absorbed (scaled to counter the effect of expansion; see text), of the gas temperature, and of the hydrogen ionization. In the bottom panel, we overplot the hydrogen mass fraction (dashed line).
\label{fig_ejecta_prop}
}
\end{figure}

Figure~\ref{fig_ejecta_prop} shows the evolution of some ejecta properties from 350\,d until 1000\,d after explosion  in the model s15p2Pwr1e40 (i.e., the model that includes a shock power of $10^{40}$\,\ergs). In the top panel, we show the total power absorbed in the ejecta due to both radioactive decay (which is biased toward the inner, denser, ejecta layers where \nifs\ is present), and shock power (which by design is deposited exclusively within the outer dense shell at 8000\,\kms). The curves are scaled by the quantity $(t_n/t_0)^3$ to counter the effect of expansion, where $t_0=350$\,d is the initial time in the sequence and $t_n$ corresponds to one of the $n$ epochs spanning from $t_0$ until 1000\,d.  The offset between the curves arises from the drop in decay power and the increasing $\gamma$-ray escape with time: 87\% of the emitted decay power is trapped at 350\,d and this drops to only 26\% at 1000\,d. The shock power is chosen constant so all curves overlap in the outer region. Evidently, the shock power reaches a maximum value in the outer regions that rivals the maximum decay power absorbed in the inner regions. As time progresses, because of the exponential decline of radioactive decay, the shock power becomes the dominant power source. This is also seen in the middle panel showing the gas temperature, which systematically decreases with time in the inner ejecta layers while it remains roughly constant in some parts of the dense shell centered at 8000\,\kms. In the intermediate regions between 4000 and 8000\,\kms, the gas temperature is boosted by the absorption of UV radiation emitted from the dense shell. In the absence of shock power, the temperature beyond a few 1000\,\kms\ drops to around 1000\,K (not shown), while with shock power it reaches a few 1000\,K in these intermediate regions, and peaks at 15,000\,K in the dense shell. In the inner ejecta regions, the evolution follows what was described for earlier times by \citet{DH20_shuffle} and \citet{D21_sn2p_neb} in similar shuffled-shell models of the explosion of a 15\,\msun\ star. In the inner ejecta, the temperature and ionization (bottom panel of Fig.~\ref{fig_ejecta_prop}) exhibit rapid variations because of the strong  chemical stratification (this contrasts with the smooth profile of the decay power absorbed).  In the outer ejecta, the hydrogen ionization is relatively  high except in the central, clumped part of the dense shell where the recombination rates are boosted (see, for example, discussion for the effect of clumping in Type II SN ejecta by \citealt{d18_fcl}). The oxygen ionization exhibits a similar profile (hence not shown), being low in the O-rich zones, and with O$^+$ dominating within the dense shell.

\begin{figure}[th]
\centering
\includegraphics[width=\hsize]{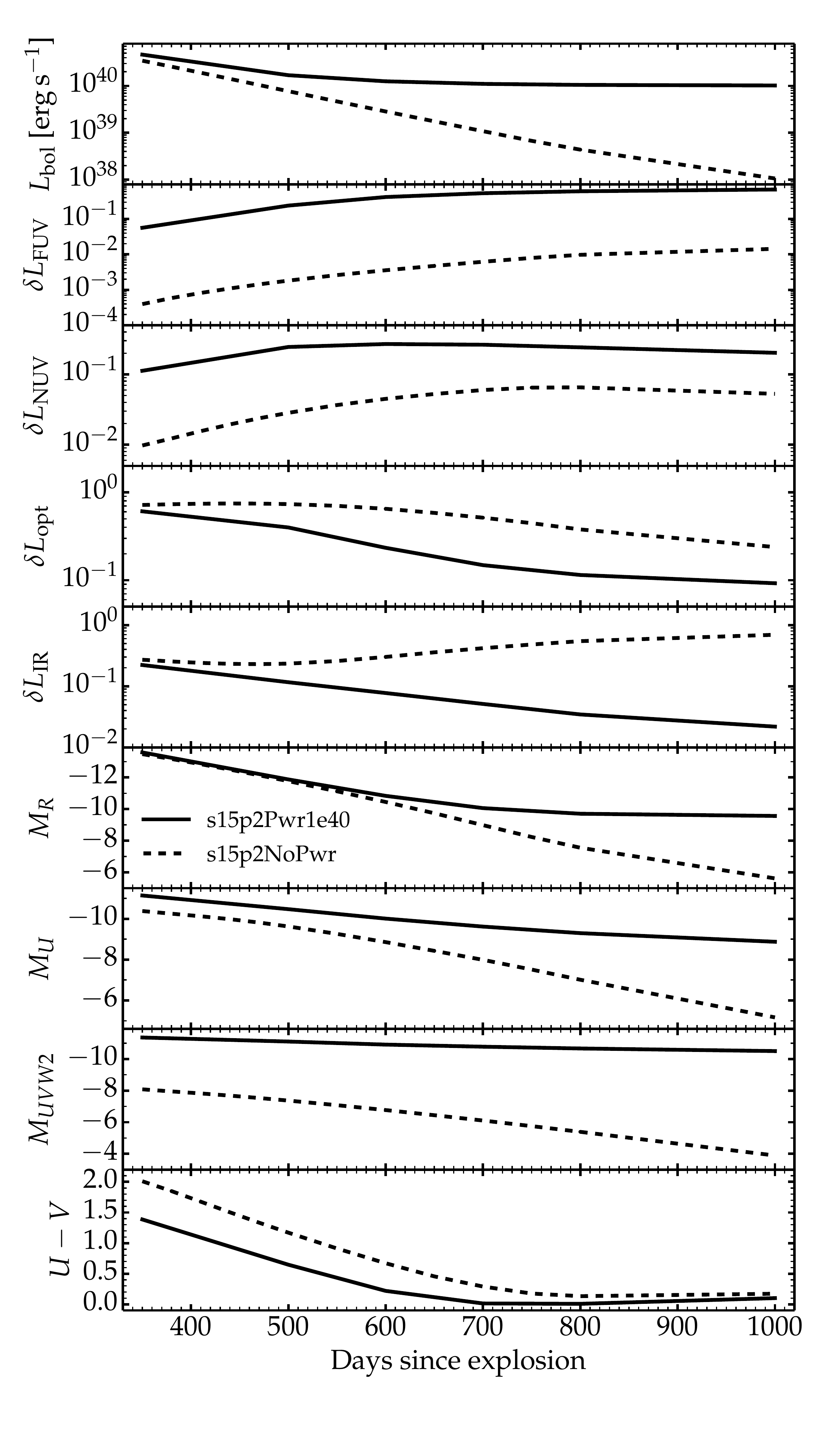}
\vspace{-1.2cm}
\caption{Evolution of photometric properties at late times for models s15p2Pwr1e40 (solid) and s15p2NoPwr (dashed). From top to bottom, we show the bolometric luminosity, the fractional power emerging in the far-UV ($\delta L_{\rm FUV}$; $1000 < \lambda/\AA < 2000$), the near-UV ($\delta L_{\rm NUV}$; $2000 < \lambda/\AA < 3500$), the optical ($\delta L_{\rm opt}$; $3500 < \lambda/\AA < 9500$), the IR ($\delta L_{\rm IR}$; $\lambda/\AA > 9500$), followed by the $R$-, $U$-, and $UVW2$-band magnitudes and the $U-V$ color.
\label{fig_phot}
}
\end{figure}

\begin{figure*}[h]
\centering
\includegraphics[width=0.49\hsize]{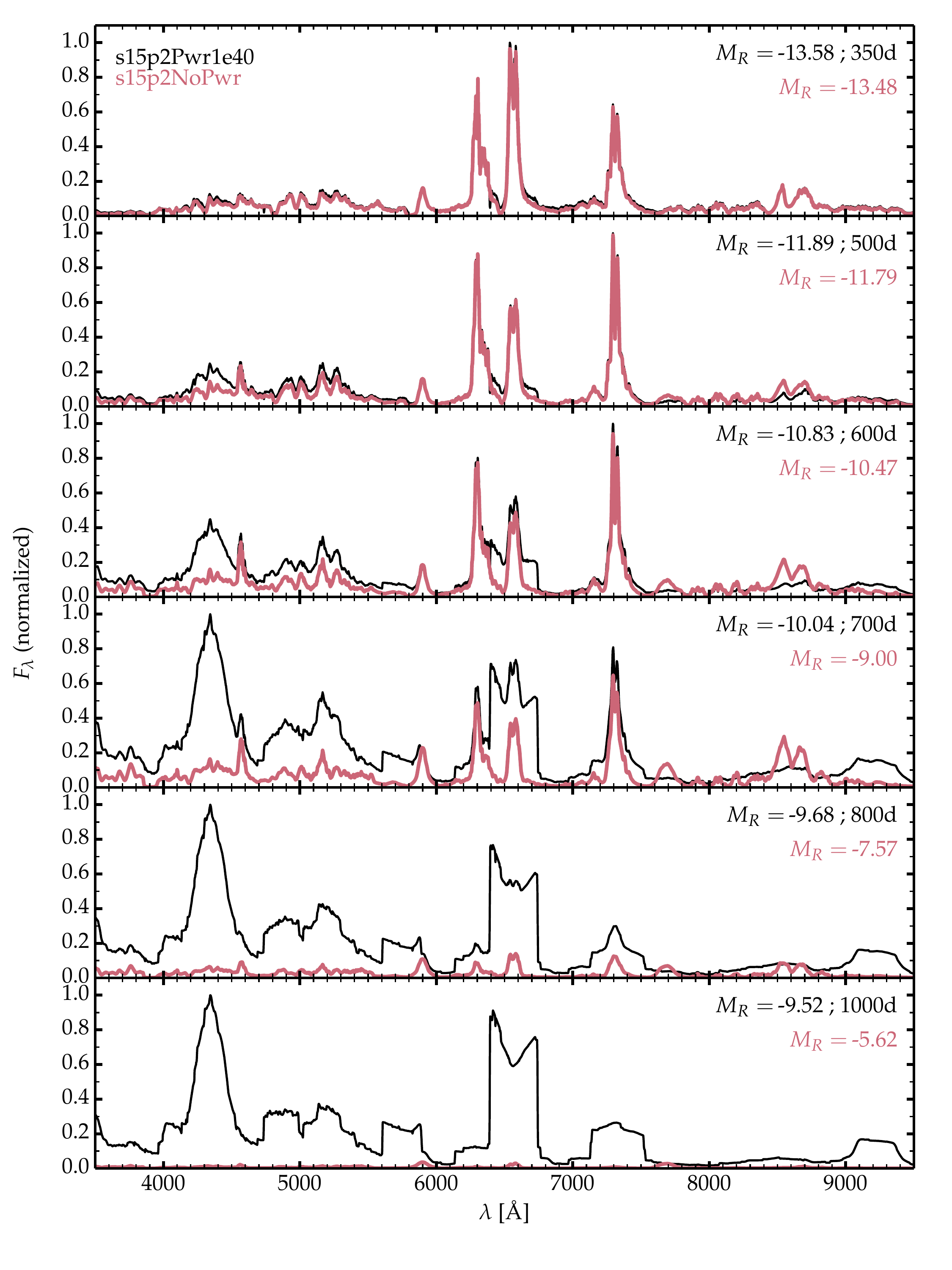}
\includegraphics[width=0.49\hsize]{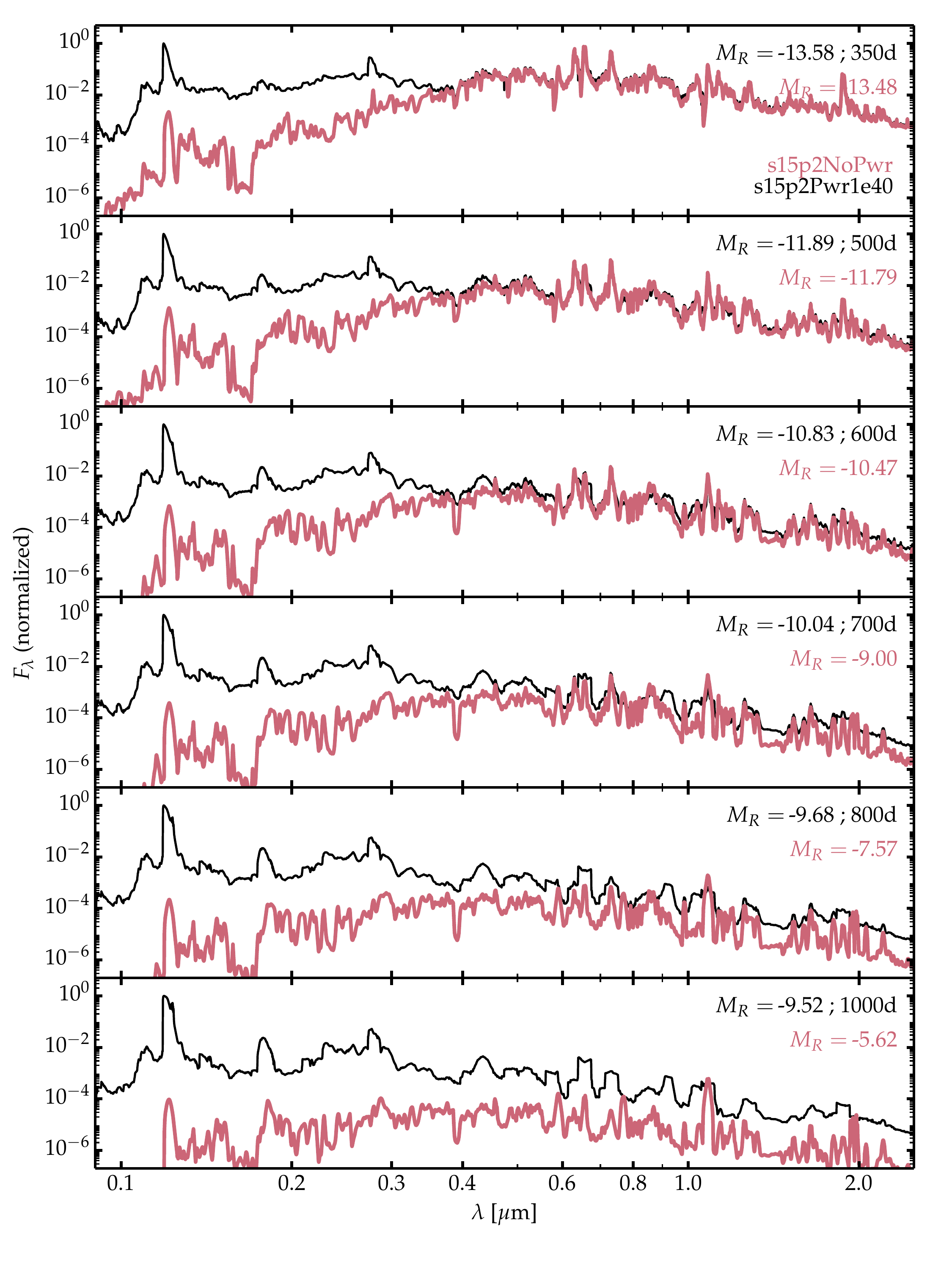}
\vspace{-0.4cm}
\caption{Comparison of multiepoch, normalized spectra from 350\,d to 1000\,d between models s15p2NoPwr (red) and s15p2Pwr1e40 (black) over the optical range (left) and over the UV, optical, and near-IR ranges (right; flux density is shown on a log scale). The same normalization factor is applied to both models at each epoch in order to preserve the relative offset between the spectra.
\label{fig_s15p2_evol}
}
\end{figure*}

\subsection{Photometric properties}

Figure~\ref{fig_phot} illustrates the photometric evolution of model s15p2 with and without shock power. As discussed by \citet{DH_interaction_22}, the impact of a shock power of $10^{40}$\,\ergs\ at 350\,d is modest in the optical, IR, and bolometric luminosity, but it is clearly visible in the $U$ band and even more so in the UV. As time progresses beyond 350\,d, the contribution from radioactive decay decreases and the contribution from shock power grows, dominating totally at 1000\,d. In the optical (e.g., the $R$ band; 6th panel from top) this switch from decay-powered to shock-powered appears as a knee in the light curve at $\sim 600$--700\,d, about 100--200\,d later than it does in $L_{\rm bol}$. This knee appears in spite of the constancy of the shock power that is introduced. In other words, the knee results from the changing contrast between shock power and decay power, driven here by the intrinsic, exponential decline of radioactive decay. In the model without shock power, the bulk of the flux emerges in the optical (about 70\%) and in the IR (about 25\%) at 350\,d, but in the model with shock power, the flux emerges primarily in the UV. As time passes, the model without (with) power becomes even more biased in favor of the IR (UV). As predicted already at earlier times, the UV range is the spectral region to investigate to capture the most obvious signatures of interaction, where the model with shock power remains 3--5\,mag brighter than its counterpart without shock power. As advocated by \citet{D22_lsst} and confirmed here for SNe~II 1--3\,yr after explosion, the $U$-band survey of the Vera Rubin Observatory should detect SNe II with an ``anomalously'' high $U$-band brightness (for the present, mostly conceptual study, the discussion holds for filters having a similar passband, such as for example the $u$-band SDSS filter and the $U$-band Johnson filter).

\subsection{Spectroscopic properties}

The spectral evolution of models s15p2NoPwr and s15p2Pwr1e40 is shown in Fig.~\ref{fig_s15p2_evol}, for both the optical (left column) and the full UV, optical, and near-IR ranges (right column; contributions in the optical range from individual species are illustrated in the Appendix, in Fig. \ref{fig_spec_ions}). This figure gives some clues about the photometric properties shown in Fig.~\ref{fig_phot}. In the optical, the main feature introduced by the shock power at 8000\,\kms\ is the excess H$\alpha$, optically thin emission, producing a broad boxy profile that contrasts with the narrower H$\alpha$ component  powered by radioactive decay and associated with hydrogen advected inward in the inner, slow-moving, ejecta. Hence, apart from this weak broad boxy H$\alpha$ emission, the entire optical spectrum is powered by radioactive decay (this is not the case for the UV -- see below). As time passes, the model powered exclusively by radioactive decay becomes increasingly faint. In contrast, in the model with shock power, we see the morphing of the ejecta from being radioactively powered and dominated by narrower lines forming in the inner ejecta, to being shock powered and dominated by broader lines forming in the outermost layers of the ejecta (in the dense shell). At intermediate epochs (e.g., $\sim 700$\,d), numerous lines show a hybrid morphology with both a narrow and a broad component (e.g., H$\alpha$ and \oidoub), while others exhibit only a narrow component (e.g., \mgi\ and \caiidoub). Here, the identification of the contribution from the outer dense shell powered by the shock is obvious because we assume spherical symmetry. If the dense shell were asymmetric, or if it is were spherical but rammed into an asymmetric CSM, the emission would no longer produce broad, boxy, symmetric line profiles but instead narrower, or skewed, or spiky line profiles. Examples such as SN\,1993J with its broad, boxy, symmetric line profiles at 976\,d \citep{matheson_93j_00a} suggest that in some cases, the outer dense shell and the CSM are probably roughly spherical on large scales.

At the latest epoch in model s15p2Pwr1e40, when the shock power dominates, the entire spectrum forms in the outer dense shell, whose temperature is in the range 5000--15000\,K and with a sizeable ionization (H and O are partially ionized in the dense shell). The spectrum shows strong H$\alpha$ (with a contribution from \niidoub), multiple lines of Fe\two\ below 5500\,\AA, strong \niiauroral\ (which overlaps with weak and narrow Na\one\,D emission), and broad, boxy \oidoub\ which overlaps with and causes a blue hump on H$\alpha$. Further to the red, \oiidoub\ overlaps with the narrow \caiidoub\ arising from the inner ejecta and produces a broad, boxy, symmetric profile with a narrow central bump. A similar profile is obtained for He\one\,10,830\,\AA. The model  predicts a strong broad Mg\two\,$\lambda\lambda$\,9218, 9244 line flux, although the line appears with slanted wings (in part because it overlaps with some background Fe\two\ emission).

 \begin{figure*}[h]
\centering
\includegraphics[width=0.8\hsize]{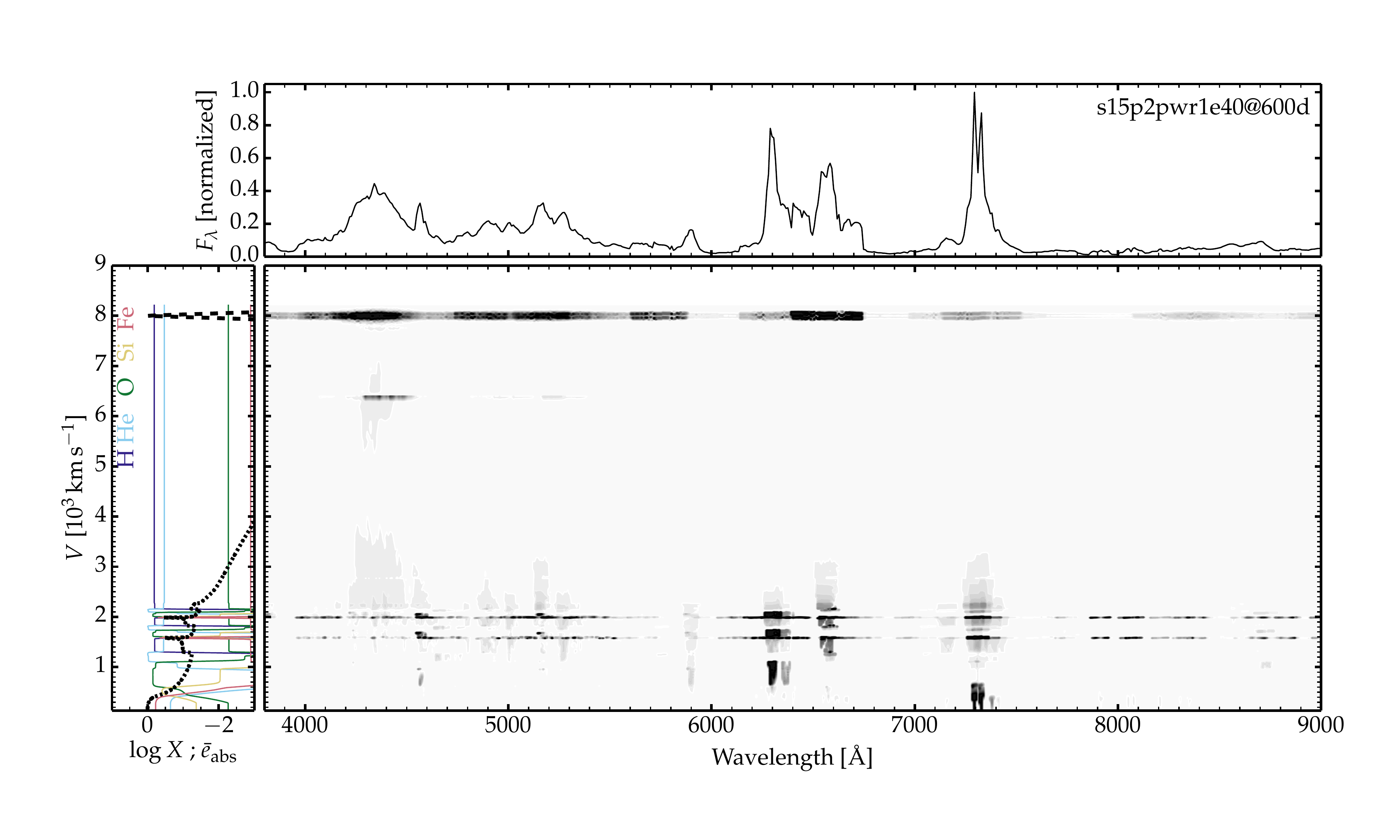}
\vspace{-0.4cm}
\caption{Illustration of the spatial regions (here shown in velocity space) contributing to the emergent flux in
model s15p2Pwr1e40. The main panel shows the observer's frame luminosity contribution $\delta L_{\lambda,R}$ versus wavelength and ejecta velocity. The left panel shows the composition profile versus velocity together with the power absorbed (decay power: dotted line; shock power: dashed line), while the top panel show the normalized flux density, which at each wavelength is a sum of $\delta L_{\lambda,R}$ over all ejecta layers.
\label{fig_dfr}
\label{fig}
}
\end{figure*}

The right panel in Fig.~\ref{fig_s15p2_evol} shows the full spectrum from 0.09 to 2.5\,$\mu$m. To accommodate the large variation from short to long wavelengths and between models s15p2NoPwr and s15p2Pwr1e40, the flux density is shown on a logarithmic scale. This confirms that the offset between the two models is initially confined to regions below 3000\,\AA\ (with thus a rough agreement in the $U$ band at 350\,d), but that the offset grows in time and becomes ever more pronounced at later times. More importantly, the spectrum formation regions become more and more distinct, with model  s15p2NoPwr becoming increasingly faint with a spectrum forming in the inner radiatioctive-decay powered layers, while model s15p2Pwr1e40 remains luminous with a spectrum dominated by broad lines forming in the outer ejecta. While Ly$\alpha$ and \mgiires\ are the two strongest lines in the UV range, they eventually also dominate over the UV continuum flux, especially Ly$\alpha$ which alone contains $\sim$\,65\% of the bolometric flux at 1000\,d.

   Figure~\ref{fig_dfr} illustrates the spatial origin of the emerging luminosity (or flux) in model s15p2Pwr1e40 at 600\,d after explosion. More specifically, it shows the fractional luminosity $\delta L_\lambda(R)$ at wavelength $\lambda$ arising from a narrow ejecta shell at radius $R$,
\begin{equation}
\label{eq_dfr}
\delta L_\lambda(R) = 8\pi^2\int \Delta z\
\eta(p,z,\lambda)\ e^{-\tau(p,z,\lambda)}\ p\ dp,
\end{equation}
\noindent
where $\Delta z$ is the projected shell thickness for a ray with impact parameter $p$, $\eta$ is the
emissivity along the ray at $p$ and $z$, and $\tau$ is the total ray optical depth at $\lambda$ (i.e., the integral is performed in the observer's frame) at the ejecta location $(p,z)$. Figure~\ref{fig_dfr} illustrates the location of the last interaction of photons with the ejecta material -- at such a late time this is in most lines the location where the photon was originally emitted. At this intermediate stage, the grayscale shows the hybrid formation of the spectrum with emission arising from the inner ejecta as well as the outer dense shell where shock power is deposited. This is clearly seen through the location of the emission in velocity space as well as the width of the strongest lines like H$\alpha$, which forms in two distinct velocity regions below 3000\,\kms\ and around 8000\,\kms.

\begin{table*}
\begin{center}
\caption{SN sample including references to the spectroscopic and photometric data as well as some SN characteristics.}
\label{tab_obs}
\begin{tabular}{ccccccccc}
\hline
\hline
SN                    & Redshift & Explosion date & Epochs of nebular spectra  & Photometry &    References              \\
                      &          &      (MJD)     &         (days)             &  (Band)    &                            \\
\hline
\hline
SN~2004et$^{\dagger}$ &  0.00016 &      53270.0     & 355, 464, 823, 933, 1145 &    $R$     &   (1), (2), (3), (4)       \\
SN~2007od             &  0.00578 &      54407.2$^*$ &         348, 699         &    $R$     &    (5), (6), (7)           \\
SN~2013by$^{\dagger}$ &  0.00382 &      56404.0     &         278, 457         &    $r$     &       (8), (9)             \\
SN~2013ej$^{\star}$   &  0.00219 &      56497.0     &      269, 434, 806       &    $R$     &(10), (11), (12), (13), (7) \\
SN~2014G$^{\dagger}$  &  0.1637  &      56669.6     &            342           &    $r$     &          (14)              \\
SN~2017eaw            &  0.0039  &      57886.0     &       415, 482, 900      &    $R$     &    (15), (16), (17)        \\
SN~2017ivv            &  0.003   &      58091.6     &         337, 520         &    $r$     &          (18)              \\
\hline
\hline
\end{tabular}
\begin{list}{}{}
\item $^{\dagger}$ Data downloaded from the WISeREP (\url{http://wiserep.weizmann.ac.il/home}) archive \citep{wiserep}.
\item $^{\star}$ Nebular spectra obtained from the UC Berkeley Filippenko Group's Supernova Database \citep{silverman_ucb_archive_12}: \url{http://heracles.astro.berkeley.edu/sndb/search}.
\item $^{*}$ Explosion epoch from \cite{andrews_07od_10}.
\item \textbf{References:} (1) \citet{sahu_04et_06}; (2) \citet{misra_04et_07}; (3) \citet{maguire_04et_10}; (4) \citet{faran_sn2l_14}; (5) \cite{andrews_07od_10}; (6) \citep{inserra_07od_11}; (7) \citep{dejaeger_bvri_19}; (8) \citet{valenti_13by_15}; (9) \citet{black_13by_17}; (10) \citet{valenti_13ej_14}; (11) \citet{Richmond14}; (12) \citet{bose_13ej_15}; (13) \citet{yuan_13ej_16}; (14) \citet{terreran_14G_16}; (15) \citet{buta_keel_17eaw_19}; (16) \citet{vandyk_17eaw_19}; (17) \citet{weil_17eaw_20}; (18) \citet{gutierrez_17ivv_20}.
\end{list}
\end{center}
\end{table*}

\begin{figure}
\centering
\includegraphics[width=\hsize]{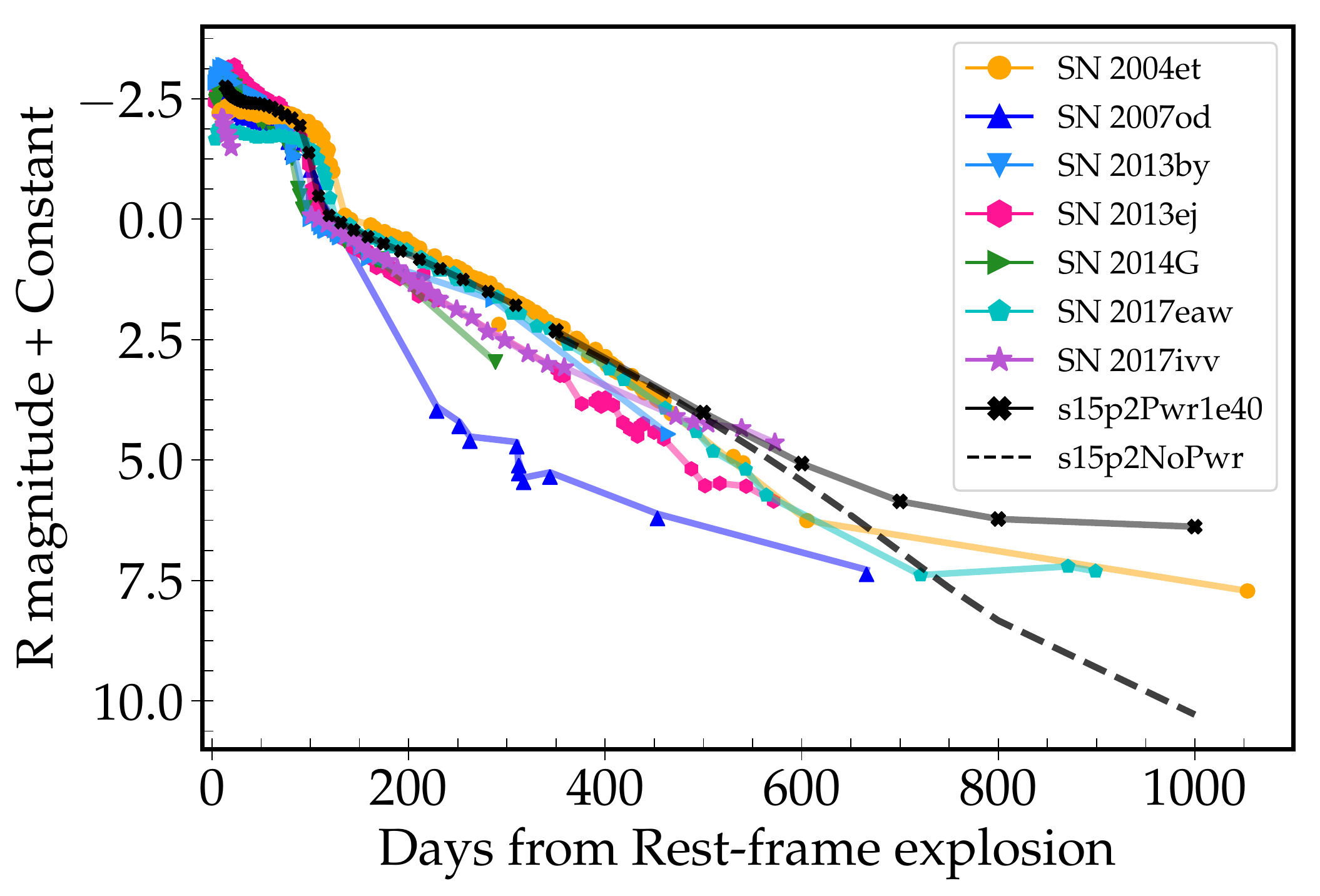}
\vspace{-0.4cm}
\caption{$R$-band light curves for model s15p2NoPwr and s15p2Pwr1e40 compared with a set of observations  (for a few objects, we show the $r$-band magnitude instead; see Table~\ref{tab_obs}). All light curves are shifted so that the magnitude is zero at the onset of the nebular phase (when the fall from the plateau stops and the SN follows the rate of \cofs\ decay).
\label{fig_lc_with_obs}
}
\end{figure}

\begin{figure}
\centering
\includegraphics[width=\hsize]{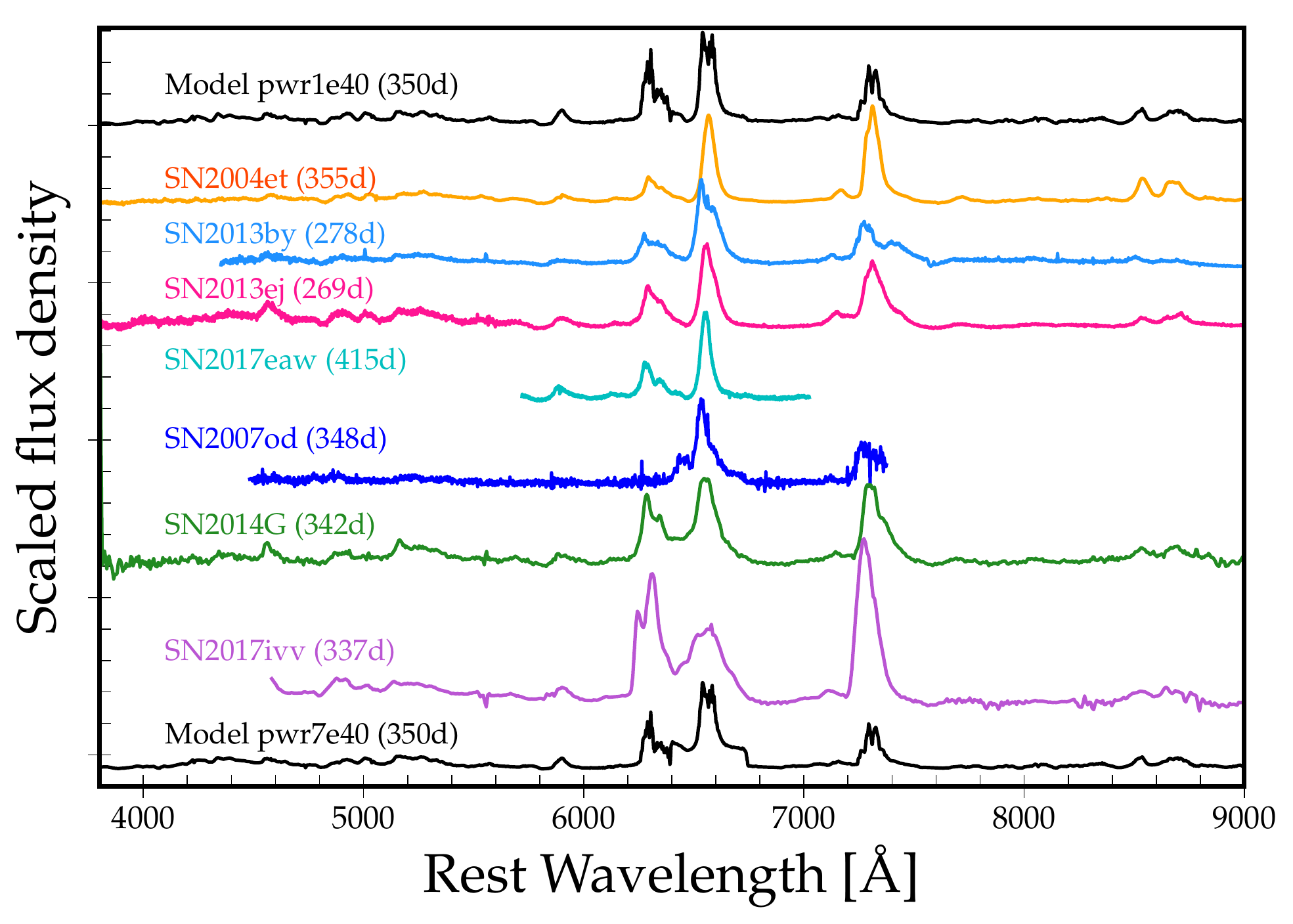}
\includegraphics[width=\hsize]{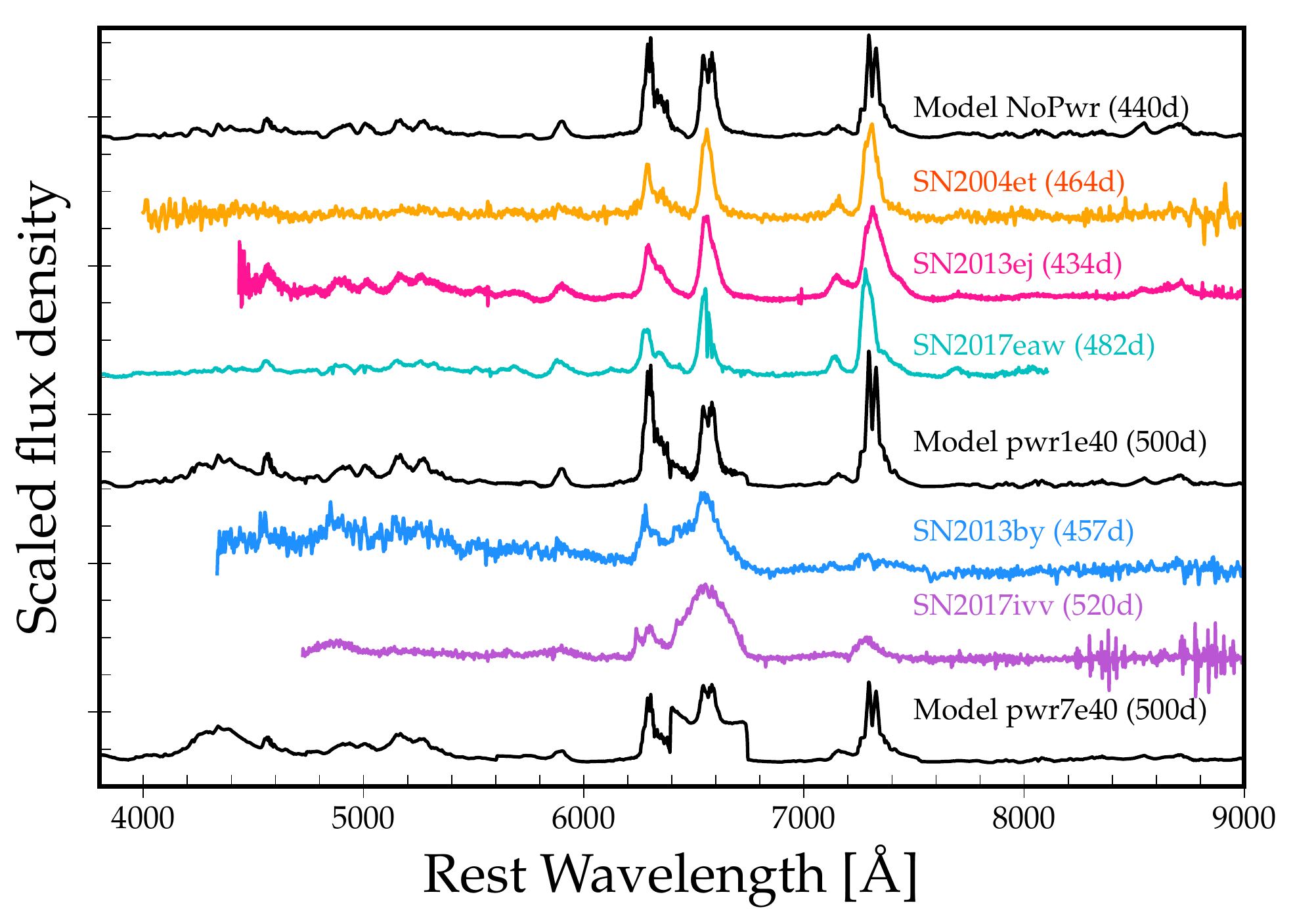}
\includegraphics[width=\hsize]{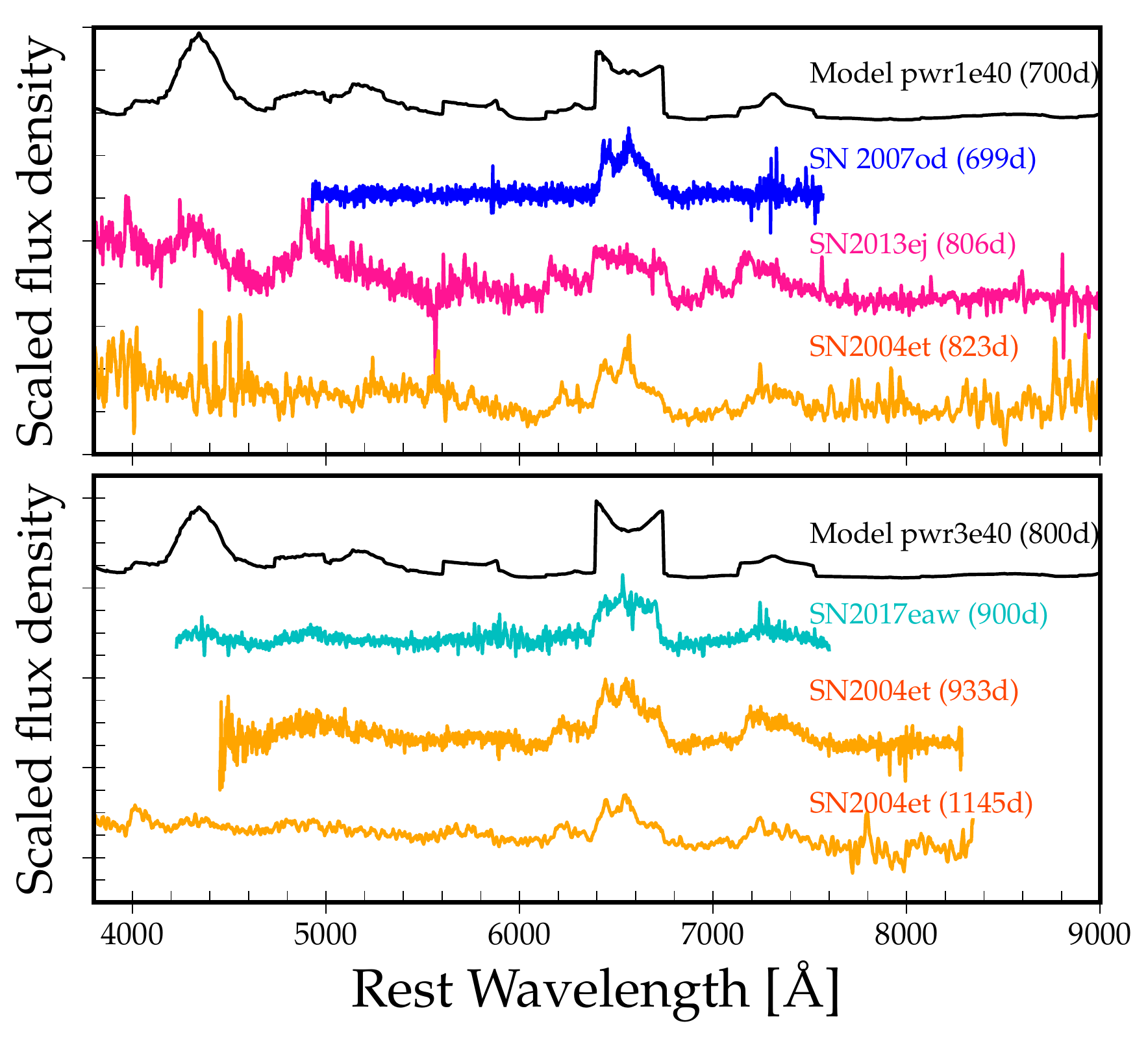}
\vspace{-0.4cm}
\caption{Spectral comparison between models and the selected sample of observations (see Table~\ref{tab_obs}) at $\sim$\,350\,d (top panel), $\sim$\,450\,d (middle panel), and 800--1000\,d (lower two panels). In each panel, we stack models and observations in strengthening signs of interaction as we progress downwards. A distinct color is used for each SN while black is used for the models (see labels; light (dark) colors correspond to the original (smoothed) spectra). The galaxy lines were removed from the observed spectra, and the data were binned to reduce the noise.
\label{fig_comp_spec_obs}
}
\end{figure}

 \begin{figure}
\centering
\includegraphics[width=\hsize]{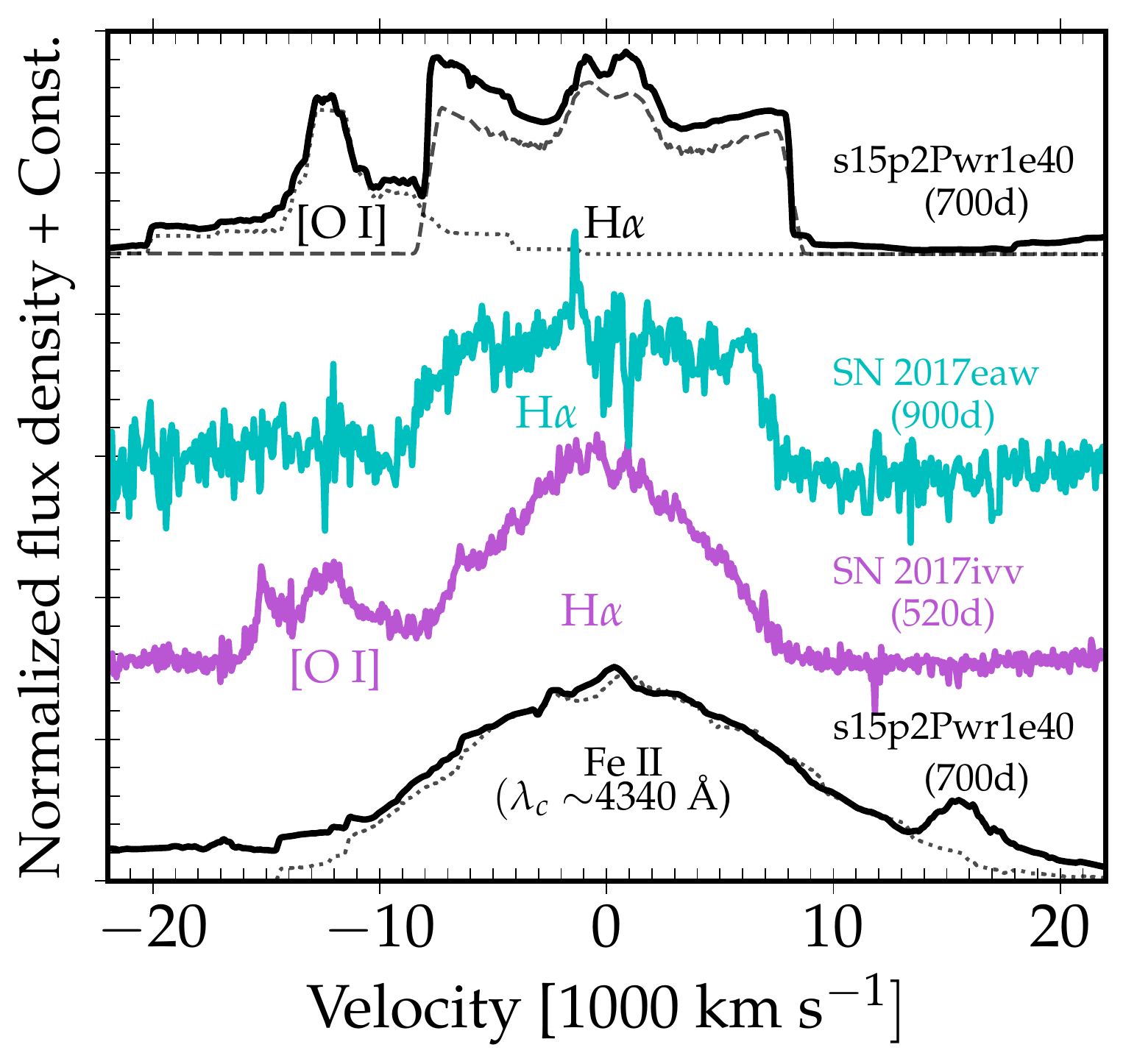}
\vspace{-0.4cm}
\caption{Illustration of some profile morphologies in model s15p2Pwr1e40 (top and bottom spectra) and the observations of SNe 2017eaw and 2017ivv. The top three curves are centered on the rest wavelength of H$\alpha$, while the bottom curve is centered on 4340\,\AA\, which corresponds to the peak emission of that Fe\two\ blend in model s15p2Pwr1e40.
\label{fig_profiles}
}
\end{figure}

\section{Comparison with observations}
\label{sect_obs}

  In this section, we compare our model results with observed Type II SNe. The sample was selected based on observed properties reminiscent of the interaction features we obtain in this work as well as on the association with ejecta/CSM interaction made in previous work. Our search was focused on objects with late-time spectra exhibiting peculiar spectral properties such as broad boxy line profiles, an anomalously strong H$\alpha$ line, and so on. The sample consists of seven objects with spectra ranging between 278 and 1145\,d from explosion, and specifically SNe 2004et, 2007od, 2013by, 2013ej, 2014G, 2017eaw, and 2017ivv. Basic information on each object and relevant references are presented in Table~\ref{tab_obs}.

Figure~\ref{fig_lc_with_obs} compares the $R$-band light curve of the sample of observations with the models presented in Section~\ref{sect_res}. To facilitate the comparison, the light curves have been shifted so that the magnitude is zero at the onset of the nebular phase (i.e., when the magnitude falls on a slow linear decline after the abrupt drop at the end of the high-brightness, plateau phase). This corrects for differences in distance, reddening, and \nifs\ mass. The element of interest is to identify if the $R$-band magnitude follows continuously a roughly linear decline\footnote{The decline may not be exactly linear, but should be close, if the $R$-band magnitude is only affected by a progressive and weak color shift and enhanced $\gamma$-ray escape.} like in model s15p2NoPwr or if a kink appears in the light curve like in model s15p2Pwr1e40. SNe 2013by, 2013ej, and 2014G do not exhibit any $R$-band light-curve kink. Their light curves may not extend sufficiently late to make a strong case, but it implies nonetheless that interaction is at most a weak component to the flux in the $R$ band. In contrast, SNe 2004et, 2017eaw, and 2017ivv exhibit an $R$-band kink at 600--1000\,d, 700--900\,d, and 350--500\,d, respectively. The exact timing is unclear because of the coarse photometric monitoring. The model s15p2Pwr1e40 with its shock power of $10^{40}$\,\ergs\ yields a qualitative agreement with the $R$-band light curves of SNe 2004et, 2017eaw, and 2017ivv. Compared to the model, the kink occurs about 200\,d earlier in SN\,2017ivv and at about the same time in SN\,2017eaw. An earlier kink implies a stronger interaction or a stronger thermalization of the shock power, or both.

Figure~\ref{fig_comp_spec_obs} shows a comparison of optical spectra for the models of Section~\ref{sect_res} and the sample of observations at about 350\,d, 450\,d, and at very late times of 800\,d to $\sim$\,1200\,d.  In each panel, we order the models and the observations such that signs of interaction  appear stronger as we progress downward. In all cases, the most obvious sign of interaction is found in the H$\alpha$ profile, with the presence of a broad but not always boxy profile on top of a narrower emission component. When the extent of this H$\alpha$ line is large, it may overlap with \oidoub\ in the blue, causing a ledge between H$\alpha$ and \oidoub\ (i.e., the flux does not go down to zero between the two lines as it typically does in the nebular-phase spectra of Type II SNe; see, for example, \citealt{leonard_99em}). At around 350\,d (top panel of Fig.~\ref{fig_comp_spec_obs}), SNe 2004et, 2013by, 2013ej, and 2017eaw show no obvious spectroscopic sign of interaction in the optical. In contrast, SNe 2014G and 2017ivv exhibit a broad H$\alpha$ component, which also boosts the red component of \oidoub. The model with a power of $7 \times 10^{40}$\,\ergs\ gives a good match to these last two SNe although the broad H$\alpha$ emission appears more boxy than in the observations.

At about 450\,d (middle panel of Fig.~\ref{fig_comp_spec_obs}), obvious signs of interaction are still absent in SNe 2004et, 2013ej, and 2017eaw, but SN\,2013by now shows a broad H$\alpha$ profile, which overlaps with \oidoub, as observed with SN\,2014G at 350\,d (top panel). Surprisingly, SN\,2013by exhibits narrow H$\alpha$ emission as well, powered by radioactive decay, but the equivalent narrow emission in  \caiidoub\ has essentially disappeared. Finally, SN\,2017ivv continues to show signs of interaction, with an H$\alpha$ line that now looks triangular, suggesting a boost of the H$\alpha$ emissivity over a range of velocities in the outer ejecta, together with emission from the inner ejecta where \oidoub\ and \caiidoub\ also form.

At times greater than 800\,d, the contribution from the inner ejecta powered by radioactive decay is becoming very small and ejecta/CSM interaction is the reason for the detection of these SNe at such late times. This explains the more boxy and broad H$\alpha$ profiles in SNe 2004et, 2007od, 2013ej, and 2017eaw. The spectra are quite noisy at those late times, but we tentatively identify broad \oidoub\ (except in SN\,2007od), whose overlap with H$\alpha$ causes a blue hump in that line. A similar overlap may be at the origin of the skewed and slanted feature at 7300\,\AA\ in SN\,2013ej at 806\,d. This line may be a composite of \caiidoub\ and \oiidoub, and is seen as a broad boxy profile in SN\,2004et at 933\,d. In the blue part of the spectrum, H$\beta$ as well as multiple Fe\two\ lines (all broad) are likely present, as in model s15p2Pwr1e40. The narrow features that appear in numerous line profiles are unlikely to arise from unshocked CSM or from the inner ejecta. They probably arise instead from the small overlap between broad lines, or from the clumpy structure of the interaction region (see, for example, discussion of SN\,1993J by \citealt{matheson_93j_00b}).

The model s15p2Pwr1e40 gives a satisfactory match to the observations of SNe 2004et and 2017eaw at the latest time (bottom panel of Fig.~\ref{fig_comp_spec_obs}), although the H$\alpha$ line profile exhibits a central depression that is not seen in any of the observations. This dip is likely caused by an optical depth effect arising from the adoption of a narrow dense shell, causing enhanced absorption at the limb (this effect is similar to limb darkening). Breaking the lateral coherence of the shell would enhance photon escape for regions moving perpendicular to the line of sight, and would thus likely erase this central dip (see discussion of a similar effect in the context of clumping in Wolf-Rayet winds by \citealt{flores_shell_22}). Some emission features, for example around 4340\,\AA, are clearly not boxy but triangular. This morphology probably arises from the overlap of many emission features because the emission centered at 4340\,\AA\ is much broader than H$\alpha$ (Fig~\ref{fig_profiles}). Emission over a range of velocities would also tend to produce a more triangular (or Gaussian) profile shape (though potentially still with a flat top), while our models tend to emit exclusively from within the narrow dense shell at very late times.

\section{Discussion}
\label{sect_disc}

The main uncertainty in this work is the treatment of shock power in \cmfgen. As discussed by \citet{DH_interaction_22}, performing one-dimensional radiation hydrodynamics of the interaction at late times is superfluous since conditions are optically thin and the dynamical effects are vanishingly small, while the treatment of the gas in LTE is inappropriate (see discussion in \citealt{D15_2n}). Since the power released by the shock between ejecta and CSM is known analytically, one may directly model the nonLTE problem within a radiative transfer code like \cmfgen, as done by \citet{DH_interaction_22}.  The benefits are enormous, as shown with the simulations presented here, but there are numerous uncertainties that impact the results quantitatively. While the shock is originally released in the form of high-energy radiation and particles, we focus here on the fraction that is absorbed by the outer ejecta. Ultimately, this fraction of thermalized X-rays requires detailed 3D multigroup radiation hydrodynamics simulations of the interaction. The thermalization also depends on the structure of the outer ejecta, and in particular whether there is a dense shell at the interface between ejecta and CSM. This dense shell may form directly at the time of shock breakout and accumulate mass as the ejecta sweeps up progressively more CSM. This shell should break up, be turbulent, and form clumps. The transfer of radiation within this structured medium is very complex. All these aspects impact the results presented here. Multiwavelength observations covering from the X-rays to the UV and optical can help constrain these structural properties.

In this and previous work, the shock power is deposited in a narrow, clumped, dense shell in the outer ejecta (here at 8000\,\kms). The high density in the shell leads to efficient reprocessing at the site of deposition, which ultimately leads to a boost in line emissivity within the shell, causing a broad boxy profile with vertical edges in numerous emission lines like H$\alpha$. That is, all the emission comes from this narrow shell at late times. In practice, if the dense shell were less dense and massive, some of the shock power deposited would be reprocessed deeper in the ejecta and would boost the line emissivity over a range of velocities (i.e., ejecta depth), which would produce flat-top profiles with extended slanted wings (rather than a profile with a sharp, vertical edge).

This ``distributed" emission likely holds in SN\,2017ivv, where one cannot distinguish the broad from the narrow emission component at very late times since the H$\alpha$ profile is smooth, triangular with no sharp jump. At earlier times, the H$\alpha$ profile is dominated by a narrow component, which requires the presence of hydrogen at low velocity in the ejecta. This is in tension with the Type IIb classification proposed by \citet{gutierrez_17ivv_20}, who also suggested a peculiar formation process of \oidoub\ to explain the near equal intensity in each component of the doublet. The peculiar intensity ratio is explained here as the overlap of the red component of \oidoub\ with the blue wing of the broad H$\alpha$ component. Finally, it is intriguing that SN\,2017ivv exploded in one of the lowest luminosity host galaxies, suggestive of a low metallicity. Such conditions are unfavorable for a high RSG mass-loss rate. Perhaps the progenitor of SN\,2017ivv was in an interacting binary and was undergoing mass transfer in the last decades or centuries prior to core collapse.

For simplicity, all simulations in this work assumed spherical symmetry. This is a sensible first step since spherical symmetry is more uniquely defined than asymmetry, which can take an infinite number of forms. Nonetheless, the interaction between ejecta and CSM should be clumped on small scales, and could even exhibit asymmetries on large scales because of the asymmetry of the ejecta, the CSM, or both. Small-scale clumping would lead to small-scale variations within line profiles (as observed in SN\,1993J; \citealt{matheson_93j_00b}). Large-scale asymmetry would produce asymmetric, skewed rather than boxy symmetric line profiles. Such boxy symmetric profiles are seen for example in SN\,1993J, which clearly demonstrate that asymmetry is not universal in core-collapse SNe and may in fact be mostly present deeper in the ejecta (the radio emission from SN\,1993J two years after explosion is also largely spherical; see, for example, \citealt{marcaide_93J_95}).

A related issue is the prediction by the models at late times of an H$\alpha$ emission profile that exhibits a central dip and two horns rather than a flat top. This effect arises because the line optical depth is large, causing an enhanced attenuation of the photons emitted from the ejecta regions moving perpendicular to the line of sight. The lack of such a dip in the observations argues from the break-up of the dense shell at the ejecta/CSM interface, as advocated in the context of clumpy Wolf-Rayet winds \citep{flores_shell_22}. Similar simulations in the context of SNe are straightforward and will be explored in the near future. A ramification of this feature is that in some cases where the dense shell would retain a full lateral coherence without breaking up (if it retained a spherical geometry), the H$\alpha$ emission profile should appear double peaked. Interestingly, double-peaked profile are sometimes observed in SNe~II at late times (e.g., iPTF14hls; \citealt{andrews_smith_iptf14hls_18}) and generally attributed to disk-like geometries. An interesting alternative is that the double-peaked profile could stem from an optical-depth effect, similar to limb darkening, in an emitting narrow dense shell.

The possibility for late-time interaction in core-collapse SNe has implications for the post-explosion photometry at the site of the explosion. This is generally used to argue for the disappearance of the progenitor, but often a source of emission is found and it is not clear whether one observes a binary companion, a cluster of stars \citep{maund_prog_15,sun_prog_21}, or even a light echo \citep[e.g.,][]{schmidt_91T_94}. It may instead be that the emitting source arises from ejecta/CSM interaction (see, for example, \citealt{rizzo_late_sn2p_23} or \citealt{vandyk_prog_23}). Observing core-collapse SNe at late times in the $U$ band together with the optical would help determine the properties of the progenitor site and whether there is ongoing ejecta/CSM interaction. This could be performed in part with the Vera Rubin Observatory, which will collect deep photometric data for a huge number of core-collapse SNe. However, much more compelling evidence for interaction would be given by UV observations for example with the {\it Hubble Space Telescope} or possibly in the more distant future with UVEX \citep{uvex}.

\section{Conclusions}
\label{sect_conc}

In this work, we have presented a set of simulations for Type II SN ejecta from 350\,d until 1000\,d after explosion. Chemical mixing in the metal-rich regions was handled with the shuffled-shell approach of \citet{DH20_shuffle} to enforce macroscopic mixing without introducing microscopic mixing. This allows the deposition of decay power throughout the inner ejecta, powering the metal lines observed in SNe~II at nebular phases. One model (named s15p2NoPwr) was evolved until 1000\,d with the time-dependent solver, a time step set to 10\% of the current time, and under the influence of radioactive-decay heating from \nifs\ and \cofs. In another model (named s15p2Pwr1e40), we injected in addition a constant power of $10^{40}$\,\ergs\ in a narrow dense shell of 0.1\,\msun\ located at 8000\,\kms.  Molecule formation and dust formation in the ejecta were ignored.  The results are qualitatively analogous to a similar model presented by \citet{DH_interaction_22} but quantitatively widely different.

From 350\,d to 1000\,d, model s15p2Pwr40 morphed from being decay powered to being interaction powered, while exhibiting hybrid properties in between. In the optical, the evidence for this transformation is best apparent from the line profiles of H$\alpha$ or \oidoub. Originally forming in the slow, decay-powered layers of the ejecta, a broad component appears which eventually dominates at late times. In our spherical model, this broad component is boxy and flat topped, and can give rise to overlap with other lines (e.g., the blue edge of H$\alpha$ overlapping with the red component of \oidoub). At late times, all lines are broad and boxy and can lead to strong overlap (there is one exception to this with the strong broad triangular feature around 4300\,\AA, which is a blend of Fe\two\ transitions). In the UV, the impact of interaction is large since the bulk of the emission is channeled in that UV region, and most strongly in Ly$\alpha$ and \mgiires. Both lines are strong, optically thick, and skewed. The rest of the UV flux is primarily associated with emission from a forest of Fe\two\ transitions, including the strong feature around 1800\,\AA.

Compared with observations, our model s15p2Pwr1e40 yields a good match to a sample of SNe II with broad and strong H$\alpha$ emission at late times, namely SNe 2004et, 2007od, 2013by, 2013ej, 2014J, 2017eaw, and 2017ivv. All show similar behavior, but the strengthening of the H$\alpha$ line turns on at distinct post-explosion times, around 200\,d in SN\,2017ivv but only after 600\,d in SN2017eaw. This strengthening of the H$\alpha$ line is in part responsible for the kink observed in the $R$ band, which is also predicted by model s15p2Pwr1e40. Although the exact level of thermalization of shock power in the outer ejecta is unknown, our results suggest that these SNe~II are ramming into a wind of modest density, probably standard for a RSG progenitor (see also \citealt{rizzo_late_sn2p_23}). Ultraviolet observations are needed to firmly establish this conclusion.

\begin{acknowledgements}

This work was supported by the ``Programme National de Physique Stellaire'' of CNRS/INSU co-funded by CEA and CNES. H.K. was funded by the Academy of Finland projects 324504 and 328898. A.V.F. is grateful for the support of the Christopher R. Redlich Fund and numerous individual donors. This work was granted access to the HPC resources of  TGCC under the 2021 and 2022 allocations A0110410554  and A0130410554  made by GENCI, France. This research has made use of NASA's Astrophysics Data System Bibliographic Services.

\end{acknowledgements}


\begin{thebibliography}{69}
\expandafter\ifx\csname natexlab\endcsname\relax\def\natexlab#1{#1}\fi

\bibitem[{{Aguilera-Dena} {et~al.}(2022){Aguilera-Dena}, {Langer},
  {Antoniadis}, {Pauli}, {Dessart}, {Vigna-G{\'o}mez}, {Gr{\"a}fener}, \&
  {Yoon}}]{aguilera_dena_ibc_22}
{Aguilera-Dena}, D.~R., {Langer}, N., {Antoniadis}, J., {et~al.} 2022, \aap,
  661, A60

\bibitem[{Andrews {et~al.}(2010)Andrews, Gallagher, Clayton, Sugerman,
  Chatelain, Clem, Welch, Barlow, Ercolano, Fabbri, Wesson, \&
  Meixner}]{andrews_07od_10}
Andrews, J.~E., Gallagher, J.~S., Clayton, G.~C., {et~al.} 2010, The
  Astrophysical Journal, 715, 541

\bibitem[{{Andrews} \& {Smith}(2018)}]{andrews_smith_iptf14hls_18}
{Andrews}, J.~E. \& {Smith}, N. 2018, \mnras, 477, 74

\bibitem[{{Andrews} {et~al.}(2011){Andrews}, {Sugerman}, {Clayton},
  {Gallagher}, {Barlow}, {Clem}, {Ercolano}, {Fabbri}, {Meixner}, {Otsuka},
  {Welch}, \& {Wesson}}]{andrews_07it_11}
{Andrews}, J.~E., {Sugerman}, B.~E.~K., {Clayton}, G.~C., {et~al.} 2011, \apj,
  731, 47

\bibitem[{{Black} {et~al.}(2017){Black}, {Milisavljevic}, {Margutti}, {Fesen},
  {Patnaude}, \& {Parker}}]{black_13by_17}
{Black}, C.~S., {Milisavljevic}, D., {Margutti}, R., {et~al.} 2017, \apj, 848,
  5

\bibitem[{{Bose} {et~al.}(2015){Bose}, {Sutaria}, {Kumar}, {Duggal}, {Misra},
  {Brown}, {Singh}, {Dwarkadas}, {York}, {Chakraborti}, {Chandola},
  {Dahlstrom}, {Ray}, \& {Safonova}}]{bose_13ej_15}
{Bose}, S., {Sutaria}, F., {Kumar}, B., {et~al.} 2015, \apj, 806, 160

\bibitem[{{Buta} \& {Keel}(2019)}]{buta_keel_17eaw_19}
{Buta}, R.~J. \& {Keel}, W.~C. 2019, \mnras, 487, 832

\bibitem[{{Chevalier}(1982{\natexlab{a}})}]{chevalier_82}
{Chevalier}, R.~A. 1982{\natexlab{a}}, \apj, 258, 790

\bibitem[{{Chevalier}(1982{\natexlab{b}})}]{chevalier_82_radio_xray}
{Chevalier}, R.~A. 1982{\natexlab{b}}, \apj, 259, 302

\bibitem[{{Chugai}(1990)}]{chugai_late_sn2p_90}
{Chugai}, N.~N. 1990, Soviet Astronomy Letters, 16, 457

\bibitem[{{Chugai} {et~al.}(2007){Chugai}, {Chevalier}, \&
  {Utrobin}}]{chugai_hv_07}
{Chugai}, N.~N., {Chevalier}, R.~A., \& {Utrobin}, V.~P. 2007, \apj, 662, 1136

\bibitem[{{de Jaeger} {et~al.}(2019){de Jaeger}, {Zheng}, {Stahl},
  {Filippenko}, {Brink}, {Bigley}, {Blanchard}, {Blanchard}, {Bradley},
  {Cargill}, {Casper}, {Cenko}, {Channa}, {Choi}, {Clubb}, {Cobb}, {Cohen}, {de
  Kouchkovsky}, {Ellison}, {Falcon}, {Fox}, {Fuller}, {Ganeshalingam}, {Gould},
  {Graham}, {Halevi}, {Hayakawa}, {Hestenes}, {Hyland}, {Jeffers}, {Joubert},
  {Kandrashoff}, {Kelly}, {Kim}, {Kim}, {Kumar}, {Leonard}, {Li}, {Lowe}, {Lu},
  {Mason}, {McAllister}, {Mauerhan}, {Modjaz}, {Molloy}, {Perley}, {Pina},
  {Poznanski}, {Ross}, {Shivvers}, {Silverman}, {Soler}, {Stegman}, {Taylor},
  {Tang}, {Wilkins}, {Wang}, {Wang}, {Yuk}, {Yunus}, \&
  {Zhang}}]{dejaeger_bvri_19}
{de Jaeger}, T., {Zheng}, W., {Stahl}, B.~E., {et~al.} 2019, \mnras, 490, 2799

\bibitem[{{Dessart} {et~al.}(2015){Dessart}, {Audit}, \& {Hillier}}]{D15_2n}
{Dessart}, L., {Audit}, E., \& {Hillier}, D.~J. 2015, \mnras, 449, 4304

\bibitem[{{Dessart} \& {Hillier}(2020{\natexlab{a}})}]{DH20_neb}
{Dessart}, L. \& {Hillier}, D.~J. 2020{\natexlab{a}}, \aap, 642, A33

\bibitem[{{Dessart} \& {Hillier}(2020{\natexlab{b}})}]{DH20_shuffle}
{Dessart}, L. \& {Hillier}, D.~J. 2020{\natexlab{b}}, \aap, 643, L13

\bibitem[{{Dessart} \& {Hillier}(2022)}]{DH_interaction_22}
{Dessart}, L. \& {Hillier}, D.~J. 2022, \aap, 660, L9

\bibitem[{{Dessart} {et~al.}(2017){Dessart}, {Hillier}, \& {Audit}}]{d17_13fs}
{Dessart}, L., {Hillier}, D.~J., \& {Audit}, E. 2017, \aap, 605, A83

\bibitem[{{Dessart} {et~al.}(2018){Dessart}, {Hillier}, \& {Wilk}}]{d18_fcl}
{Dessart}, L., {Hillier}, D.~J., \& {Wilk}, K.~D. 2018, \aap, 619, A30

\bibitem[{{Dessart} {et~al.}(2021){Dessart}, {John Hillier}, {Sukhbold},
  {Woosley}, \& {Janka}}]{D21_sn2p_neb}
{Dessart}, L., {John Hillier}, D., {Sukhbold}, T., {Woosley}, S.~E., \&
  {Janka}, H.~T. 2021, \aap, 652, A64

\bibitem[{{Dessart} {et~al.}(2022){Dessart}, {Prieto}, {Hillier},
  {Kuncarayakti}, \& {Hueichapan}}]{D22_lsst}
{Dessart}, L., {Prieto}, J.~L., {Hillier}, D.~J., {Kuncarayakti}, H., \&
  {Hueichapan}, E.~D. 2022, \aap, 666, L14

\bibitem[{{Dolan} {et~al.}(2016){Dolan}, {Mathews}, {Lam}, {Quynh Lan},
  {Herczeg}, \& {Dearborn}}]{dolan_betelgeuse_16}
{Dolan}, M.~M., {Mathews}, G.~J., {Lam}, D.~D., {et~al.} 2016, \apj, 819, 7

\bibitem[{{Eldridge} {et~al.}(2008){Eldridge}, {Izzard}, \&
  {Tout}}]{eldridge_08_bin}
{Eldridge}, J.~J., {Izzard}, R.~G., \& {Tout}, C.~A. 2008, \mnras, 384, 1109

\bibitem[{{Faran} {et~al.}(2014){Faran}, {Poznanski}, {Filippenko}, {Chornock},
  {Foley}, {Ganeshalingam}, {Leonard}, {Li}, {Modjaz}, {Serduke}, \&
  {Silverman}}]{faran_sn2l_14}
{Faran}, T., {Poznanski}, D., {Filippenko}, A.~V., {et~al.} 2014, \mnras, 445,
  554

\bibitem[{{Filippenko}(1997)}]{filippenko_rev_97}
{Filippenko}, A.~V. 1997, \araa, 35, 309

\bibitem[{{Flores} {et~al.}(2023){Flores}, {Hillier}, \&
  {Dessart}}]{flores_shell_22}
{Flores}, B.~L., {Hillier}, D.~J., \& {Dessart}, L. 2023, \mnras, 518, 5001

\bibitem[{{Fransson}(1984)}]{fransson_uv_84}
{Fransson}, C. 1984, \aap, 133, 264

\bibitem[{{Fransson} \& {Bj{\"o}rnsson}(1998)}]{fransson_bjornsson_93J_98}
{Fransson}, C. \& {Bj{\"o}rnsson}, C.-I. 1998, \apj, 509, 861

\bibitem[{{Fransson} {et~al.}(2014){Fransson}, {Ergon}, {Challis}, {Chevalier},
  {France}, {Kirshner}, {Marion}, {Milisavljevic}, {Smith}, {Bufano},
  {Friedman}, {Kangas}, {Larsson}, {Mattila}, {Benetti}, {Chornock}, {Czekala},
  {Soderberg}, \& {Sollerman}}]{fransson_10jl}
{Fransson}, C., {Ergon}, M., {Challis}, P.~J., {et~al.} 2014, \apj, 797, 118

\bibitem[{{Guti{\'e}rrez} {et~al.}(2017){Guti{\'e}rrez}, {Anderson}, {Hamuy},
  {Morrell}, {Gonz{\'a}lez-Gaitan}, {Stritzinger}, {Phillips}, {Galbany},
  {Folatelli}, {Dessart}, {Contreras}, {Della Valle}, {Freedman}, {Hsiao},
  {Krisciunas}, {Madore}, {Maza}, {Suntzeff}, {Prieto}, {Gonz{\'a}lez},
  {Cappellaro}, {Navarrete}, {Pizzella}, {Ruiz}, {Smith}, \&
  {Turatto}}]{gutierrez_pap1_17}
{Guti{\'e}rrez}, C.~P., {Anderson}, J.~P., {Hamuy}, M., {et~al.} 2017, \apj,
  850, 89

\bibitem[{{Guti{\'e}rrez} {et~al.}(2020){Guti{\'e}rrez}, {Pastorello},
  {Jerkstrand}, {Galbany}, {Sullivan}, {Anderson}, {Taubenberger},
  {Kuncarayakti}, {Gonz{\'a}lez-Gait{\'a}n}, {Wiseman}, {Inserra}, {Fraser},
  {Maguire}, {Smartt}, {M{\"u}ller-Bravo}, {Arcavi}, {Benetti}, {Bersier},
  {Bose}, {Bostroem}, {Burke}, {Chen}, {Chen}, {Della Valle}, {Dong},
  {Gal-Yam}, {Gromadzki}, {Hiramatsu}, {Holoien}, {Hosseinzadeh}, {Howell},
  {Kankare}, {Kochanek}, {McCully}, {Nicholl}, {Pignata}, {Prieto}, {Shappee},
  {Taggart}, {Tomasella}, {Valenti}, \& {Young}}]{gutierrez_17ivv_20}
{Guti{\'e}rrez}, C.~P., {Pastorello}, A., {Jerkstrand}, A., {et~al.} 2020,
  \mnras, 499, 974

\bibitem[{{Hillier} \& {Dessart}(2012)}]{HD12}
{Hillier}, D.~J. \& {Dessart}, L. 2012, \mnras, 424, 252

\bibitem[{{Immler} {et~al.}(2008){Immler}, {Modjaz}, {Landsman}, {Bufano},
  {Brown}, {Milne}, {Dessart}, {Holland}, {Koss}, {Pooley}, {Kirshner},
  {Filippenko}, {Panagia}, {Chevalier}, {Mazzali}, {Gehrels}, {Petre},
  {Burrows}, {Nousek}, {Roming}, {Pian}, {Soderberg}, \&
  {Greiner}}]{immler_06jc_08}
{Immler}, S., {Modjaz}, M., {Landsman}, W., {et~al.} 2008, \apjl, 674, L85

\bibitem[{{Inserra} {et~al.}(2011){Inserra}, {Turatto}, {Pastorello},
  {Benetti}, {Cappellaro}, {Pumo}, {Zampieri}, {Agnoletto}, {Bufano},
  {Botticella}, {Della Valle}, {Elias Rosa}, {Iijima}, {Spiro}, \&
  {Valenti}}]{inserra_07od_11}
{Inserra}, C., {Turatto}, M., {Pastorello}, A., {et~al.} 2011, \mnras, 417, 261

\bibitem[{{Jacobson-Gal{\'a}n} {et~al.}(2022){Jacobson-Gal{\'a}n}, {Dessart},
  {Jones}, {Margutti}, {Coppejans}, {Dimitriadis}, {Foley}, {Kilpatrick},
  {Matthews}, {Rest}, {Terreran}, {Aleo}, {Auchettl}, {Blanchard}, {Coulter},
  {Davis}, {de Boer}, {DeMarchi}, {Drout}, {Earl}, {Gagliano}, {Gall},
  {Hjorth}, {Huber}, {Ibik}, {Milisavljevic}, {Pan}, {Rest}, {Ridden-Harper},
  {Rojas-Bravo}, {Siebert}, {Smith}, {Taggart}, {Tinyanont}, {Wang}, \&
  {Zenati}}]{wynn_20tlf_22}
{Jacobson-Gal{\'a}n}, W.~V., {Dessart}, L., {Jones}, D.~O., {et~al.} 2022,
  \apj, 924, 15

\bibitem[{{Jerkstrand} {et~al.}(2011){Jerkstrand}, {Fransson}, \&
  {Kozma}}]{jerkstrand_87a_11}
{Jerkstrand}, A., {Fransson}, C., \& {Kozma}, C. 2011, \aap, 530, A45

\bibitem[{{Kulkarni} {et~al.}(2021){Kulkarni}, {Harrison}, {Grefenstette},
  {Earnshaw}, {Andreoni}, {Berg}, {Bloom}, {Cenko}, {Chornock}, {Christiansen},
  {Coughlin}, {Wuollet Criswell}, {Darvish}, {Das}, {De}, {Dessart}, {Dixon},
  {Dorsman}, {El-Badry}, {Evans}, {Ford}, {Fremling}, {Gansicke}, {Gezari},
  {Gotberg}, {Green}, {Graham}, {Heida}, {Ho}, {Jaodand}, {Johns-Krull},
  {Kasliwal}, {Lazzarini}, {Lu}, {Margutti}, {Martin}, {Masters}, {McKernan},
  {Nissanke}, {Parazin}, {Perley}, {Phinney}, {Piro}, {Raaijmakers},
  {Rodriguez}, {Senchyna}, {Singer}, {Spake}, {Stassun}, {Stern}, {Teplitz},
  {Weisz}, \& {Yao}}]{uvex}
{Kulkarni}, S.~R., {Harrison}, F.~A., {Grefenstette}, B.~W., {et~al.} 2021,
  arXiv:2111.15608

\bibitem[{{Langer}(2012)}]{langer_araa_mdot_12}
{Langer}, N. 2012, \araa, 50, 107

\bibitem[{{Langer} {et~al.}(1994){Langer}, {Hamann}, {Lennon}, {Najarro},
  {Pauldrach}, \& {Puls}}]{langer_massive_94}
{Langer}, N., {Hamann}, W.~R., {Lennon}, M., {et~al.} 1994, \aap, 290, 819

\bibitem[{{Leonard} {et~al.}(2000){Leonard}, {Filippenko}, {Barth}, \&
  {Matheson}}]{leonard_98S_00}
{Leonard}, D.~C., {Filippenko}, A.~V., {Barth}, A.~J., \& {Matheson}, T. 2000,
  \apj, 536, 239

\bibitem[{{Leonard} {et~al.}(2002){Leonard}, {Filippenko}, {Gates}, {Li},
  {Eastman}, {Barth}, {Bus}, {Chornock}, {Coil}, {Frink}, {Grady}, {Harris},
  {Malkan}, {Matheson}, {Quirrenbach}, \& {Treffers}}]{leonard_99em}
{Leonard}, D.~C., {Filippenko}, A.~V., {Gates}, E.~L., {et~al.} 2002, \pasp,
  114, 35

\bibitem[{{Maeder} \& {Meynet}(1987)}]{maeder_meynet_87}
{Maeder}, A. \& {Meynet}, G. 1987, \aap, 182, 243

\bibitem[{{Maguire} {et~al.}(2010){Maguire}, {Di Carlo}, {Smartt},
  {Pastorello}, {Tsvetkov}, {Benetti}, {Spiro}, {Arkharov}, {Beccari},
  {Botticella}, {Cappellaro}, {Cristallo}, {Dolci}, {Elias-Rosa}, {Fiaschi},
  {Gorshanov}, {Harutyunyan}, {Larionov}, {Navasardyan}, {Pietrinferni},
  {Raimondo}, {di Rico}, {Valenti}, {Valentini}, \&
  {Zampieri}}]{maguire_04et_10}
{Maguire}, K., {Di Carlo}, E., {Smartt}, S.~J., {et~al.} 2010, \mnras, 404, 981

\bibitem[{{Marcaide} {et~al.}(1995){Marcaide}, {Alberdi}, {Ros}, {Diamond},
  {Shapiro}, {Guirado}, {Jones}, {Krichbaum}, {Mantovani}, {Preston}, {Rius},
  {Schilizzi}, {Trigilio}, {Whitney}, \& {Witzel}}]{marcaide_93J_95}
{Marcaide}, J.~M., {Alberdi}, A., {Ros}, E., {et~al.} 1995, Science, 270, 1475

\bibitem[{{Margutti} {et~al.}(2017){Margutti}, {Kamble}, {Milisavljevic},
  {Zapartas}, {de Mink}, {Drout}, {Chornock}, {Risaliti}, {Zauderer},
  {Bietenholz}, {Cantiello}, {Chakraborti}, {Chomiuk}, {Fong}, {Grefenstette},
  {Guidorzi}, {Kirshner}, {Parrent}, {Patnaude}, {Soderberg}, {Gehrels}, \&
  {Harrison}}]{margutti_14C_16}
{Margutti}, R., {Kamble}, A., {Milisavljevic}, D., {et~al.} 2017, \apj, 835,
  140

\bibitem[{{Matheson} {et~al.}(2000{\natexlab{a}}){Matheson}, {Filippenko},
  {Barth}, {Ho}, {Leonard}, {Bershady}, {Davis}, {Finley}, {Fisher},
  {Gonz{\'a}lez}, {Hawley}, {Koo}, {Li}, {Lonsdale}, {Schlegel}, {Smith},
  {Spinrad}, \& {Wirth}}]{matheson_93j_00a}
{Matheson}, T., {Filippenko}, A.~V., {Barth}, A.~J., {et~al.}
  2000{\natexlab{a}}, \aj, 120, 1487

\bibitem[{{Matheson} {et~al.}(2000{\natexlab{b}}){Matheson}, {Filippenko},
  {Ho}, {Barth}, \& {Leonard}}]{matheson_93j_00b}
{Matheson}, T., {Filippenko}, A.~V., {Ho}, L.~C., {Barth}, A.~J., \& {Leonard},
  D.~C. 2000{\natexlab{b}}, \aj, 120, 1499

\bibitem[{{Maund} {et~al.}(2015){Maund}, {Fraser}, {Reilly}, {Ergon}, \&
  {Mattila}}]{maund_prog_15}
{Maund}, J.~R., {Fraser}, M., {Reilly}, E., {Ergon}, M., \& {Mattila}, S. 2015,
  \mnras, 447, 3207

\bibitem[{{Misra} {et~al.}(2007){Misra}, {Pooley}, {Chandra}, {Bhattacharya},
  {Ray}, {Sagar}, \& {Lewin}}]{misra_04et_07}
{Misra}, K., {Pooley}, D., {Chandra}, P., {et~al.} 2007, \mnras, 381, 280

\bibitem[{{Podsiadlowski} {et~al.}(1992){Podsiadlowski}, {Joss}, \&
  {Hsu}}]{podsiadlowski_92}
{Podsiadlowski}, P., {Joss}, P.~C., \& {Hsu}, J.~J.~L. 1992, \apj, 391, 246

\bibitem[{{Richmond}(2014)}]{Richmond14}
{Richmond}, M.~W. 2014, The Journal of the American Association of Variable
  Star Observers, 42, 333

\bibitem[{{Rizzo Smith} {et~al.}(2022){Rizzo Smith}, {Kochanek}, \&
  {Neustadt}}]{rizzo_late_sn2p_23}
{Rizzo Smith}, M., {Kochanek}, C.~S., \& {Neustadt}, J.~M.~M. 2022, arXiv
  e-prints, arXiv:2212.09763

\bibitem[{{Sahu} {et~al.}(2006){Sahu}, {Anupama}, {Srividya}, \&
  {Muneer}}]{sahu_04et_06}
{Sahu}, D.~K., {Anupama}, G.~C., {Srividya}, S., \& {Muneer}, S. 2006, \mnras,
  372, 1315

\bibitem[{{Schmidt} {et~al.}(1994){Schmidt}, {Kirshner}, {Leibundgut}, {Wells},
  {Porter}, {Ruiz-Lapuente}, {Challis}, \& {Filippenko}}]{schmidt_91T_94}
{Schmidt}, B.~P., {Kirshner}, R.~P., {Leibundgut}, B., {et~al.} 1994, \apjl,
  434, L19

\bibitem[{{Silverman} {et~al.}(2012){Silverman}, {Foley}, {Filippenko},
  {Ganeshalingam}, {Barth}, {Chornock}, {Griffith}, {Kong}, {Lee}, {Leonard},
  {Matheson}, {Miller}, {Steele}, {Barris}, {Bloom}, {Cobb}, {Coil},
  {Desroches}, {Gates}, {Ho}, {Jha}, {Kandrashoff}, {Li}, {Mandel}, {Modjaz},
  {Moore}, {Mostardi}, {Papenkova}, {Park}, {Perley}, {Poznanski}, {Reuter},
  {Scala}, {Serduke}, {Shields}, {Swift}, {Tonry}, {Van Dyk}, {Wang}, \&
  {Wong}}]{silverman_ucb_archive_12}
{Silverman}, J.~M., {Foley}, R.~J., {Filippenko}, A.~V., {et~al.} 2012, \mnras,
  425, 1789

\bibitem[{{Sukhbold} {et~al.}(2016){Sukhbold}, {Ertl}, {Woosley}, {Brown}, \&
  {Janka}}]{sukhbold_ccsn_16}
{Sukhbold}, T., {Ertl}, T., {Woosley}, S.~E., {Brown}, J.~M., \& {Janka}, H.-T.
  2016, \apj, 821, 38

\bibitem[{{Sun} {et~al.}(2021){Sun}, {Maund}, {Crowther}, {Fang}, \&
  {Zapartas}}]{sun_prog_21}
{Sun}, N.-C., {Maund}, J.~R., {Crowther}, P.~A., {Fang}, X., \& {Zapartas}, E.
  2021, \mnras, 504, 2253

\bibitem[{{Tatischeff}(2009)}]{tatischeff_93J_09}
{Tatischeff}, V. 2009, \aap, 499, 191

\bibitem[{{Terreran} {et~al.}(2016){Terreran}, {Jerkstrand}, {Benetti},
  {Smartt}, {Ochner}, {Tomasella}, {Howell}, {Morales-Garoffolo},
  {Harutyunyan}, {Kankare}, {Arcavi}, {Cappellaro}, {Elias-Rosa},
  {Hosseinzadeh}, {Kangas}, {Pastorello}, {Tartaglia}, {Turatto}, {Valenti},
  {Wiggins}, \& {Yuan}}]{terreran_14G_16}
{Terreran}, G., {Jerkstrand}, A., {Benetti}, S., {et~al.} 2016, \mnras, 462,
  137

\bibitem[{{Valenti} {et~al.}(2014){Valenti}, {Sand}, {Pastorello}, {Graham},
  {Howell}, {Parrent}, {Tomasella}, {Ochner}, {Fraser}, {Benetti}, {Yuan},
  {Smartt}, {Maund}, {Arcavi}, {Gal-Yam}, {Inserra}, \&
  {Young}}]{valenti_13ej_14}
{Valenti}, S., {Sand}, D., {Pastorello}, A., {et~al.} 2014, \mnras, 438, L101

\bibitem[{{Valenti} {et~al.}(2015){Valenti}, {Sand}, {Stritzinger}, {Howell},
  {Arcavi}, {McCully}, {Childress}, {Hsiao}, {Contreras}, {Morrell},
  {Phillips}, {Gromadzki}, {Kirshner}, \& {Marion}}]{valenti_13by_15}
{Valenti}, S., {Sand}, D., {Stritzinger}, M., {et~al.} 2015, \mnras, 448, 2608

\bibitem[{{Van Dyk} {et~al.}(2023){Van Dyk}, {de Graw}, {Baer-Way}, {Zheng},
  {Filippenko}, {Fox}, {Smith}, {Brink}, {de Jaeger}, {Kelly}, \&
  {Vasylyev}}]{vandyk_prog_23}
{Van Dyk}, S.~D., {de Graw}, A., {Baer-Way}, R., {et~al.} 2023, \mnras, 519,
  471

\bibitem[{{Van Dyk} {et~al.}(2019){Van Dyk}, {Zheng}, {Maund}, {Brink},
  {Srinivasan}, {Andrews}, {Smith}, {Leonard}, {Morozova}, {Filippenko},
  {Conner}, {Milisavljevic}, {de Jaeger}, {Long}, {Isaacson}, {Crossfield},
  {Kosiarek}, {Howard}, {Fox}, {Kelly}, {Piro}, {Littlefair}, {Dhillon},
  {Wilson}, {Butterley}, {Yunus}, {Channa}, {Jeffers}, {Falcon}, {Ross},
  {Hestenes}, {Stegman}, {Zhang}, \& {Kumar}}]{vandyk_17eaw_19}
{Van Dyk}, S.~D., {Zheng}, W., {Maund}, J.~R., {et~al.} 2019, \apj, 875, 136

\bibitem[{{Weil} {et~al.}(2020){Weil}, {Fesen}, {Patnaude}, \&
  {Milisavljevic}}]{weil_17eaw_20}
{Weil}, K.~E., {Fesen}, R.~A., {Patnaude}, D.~J., \& {Milisavljevic}, D. 2020,
  \apj, 900, 11

\bibitem[{{Weiler} {et~al.}(1991){Weiler}, {van Dyk}, {Discenna}, {Panagia}, \&
  {Sramek}}]{weiler_79c_91}
{Weiler}, K.~W., {van Dyk}, S.~D., {Discenna}, J.~L., {Panagia}, N., \&
  {Sramek}, R.~A. 1991, \apj, 380, 161

\bibitem[{{Wellstein} \& {Langer}(1999)}]{wellstein_langer_99}
{Wellstein}, S. \& {Langer}, N. 1999, \aap, 350, 148

\bibitem[{{Yaron} \& {Gal-Yam}(2012)}]{wiserep}
{Yaron}, O. \& {Gal-Yam}, A. 2012, \pasp, 124, 668

\bibitem[{{Yaron} {et~al.}(2017){Yaron}, {Perley}, {Gal-Yam}, {Groh}, {Horesh},
  {Ofek}, {Kulkarni}, {Sollerman}, {Fransson}, {Rubin}, {Szabo}, {Sapir},
  {Taddia}, {Cenko}, {Valenti}, {Arcavi}, {Howell}, {Kasliwal}, {Vreeswijk},
  {Khazov}, {Fox}, {Cao}, {Gnat}, {Kelly}, {Nugent}, {Filippenko}, {Laher},
  {Wozniak}, {Lee}, {Rebbapragada}, {Maguire}, {Sullivan}, \&
  {Soumagnac}}]{yaron_13fs_17}
{Yaron}, O., {Perley}, D.~A., {Gal-Yam}, A., {et~al.} 2017, Nature Physics, 13,
  510

\bibitem[{{Yoon}(2017)}]{yoon_wr_17}
{Yoon}, S.-C. 2017, \mnras, 470, 3970

\bibitem[{{Yuan} {et~al.}(2016){Yuan}, {Jerkstrand}, {Valenti}, {Sollerman},
  {Seitenzahl}, {Pastorello}, {Schulze}, {Chen}, {Childress}, {Fraser},
  {Fremling}, {Kotak}, {Ruiter}, {Schmidt}, {Smartt}, {Taddia}, {Terreran},
  {Tucker}, {Barbarino}, {Benetti}, {Elias-Rosa}, {Gal-Yam}, {Howell},
  {Inserra}, {Kankare}, {Lee}, {Li}, {Maguire}, {Margheim}, {Mehner}, {Ochner},
  {Sullivan}, {Tomasella}, \& {Young}}]{yuan_13ej_16}
{Yuan}, F., {Jerkstrand}, A., {Valenti}, S., {et~al.} 2016, \mnras, 461, 2003

\end{thebibliography}

\appendix

\section{Additional figures}

\begin{figure}
\centering
\includegraphics[width=\hsize]{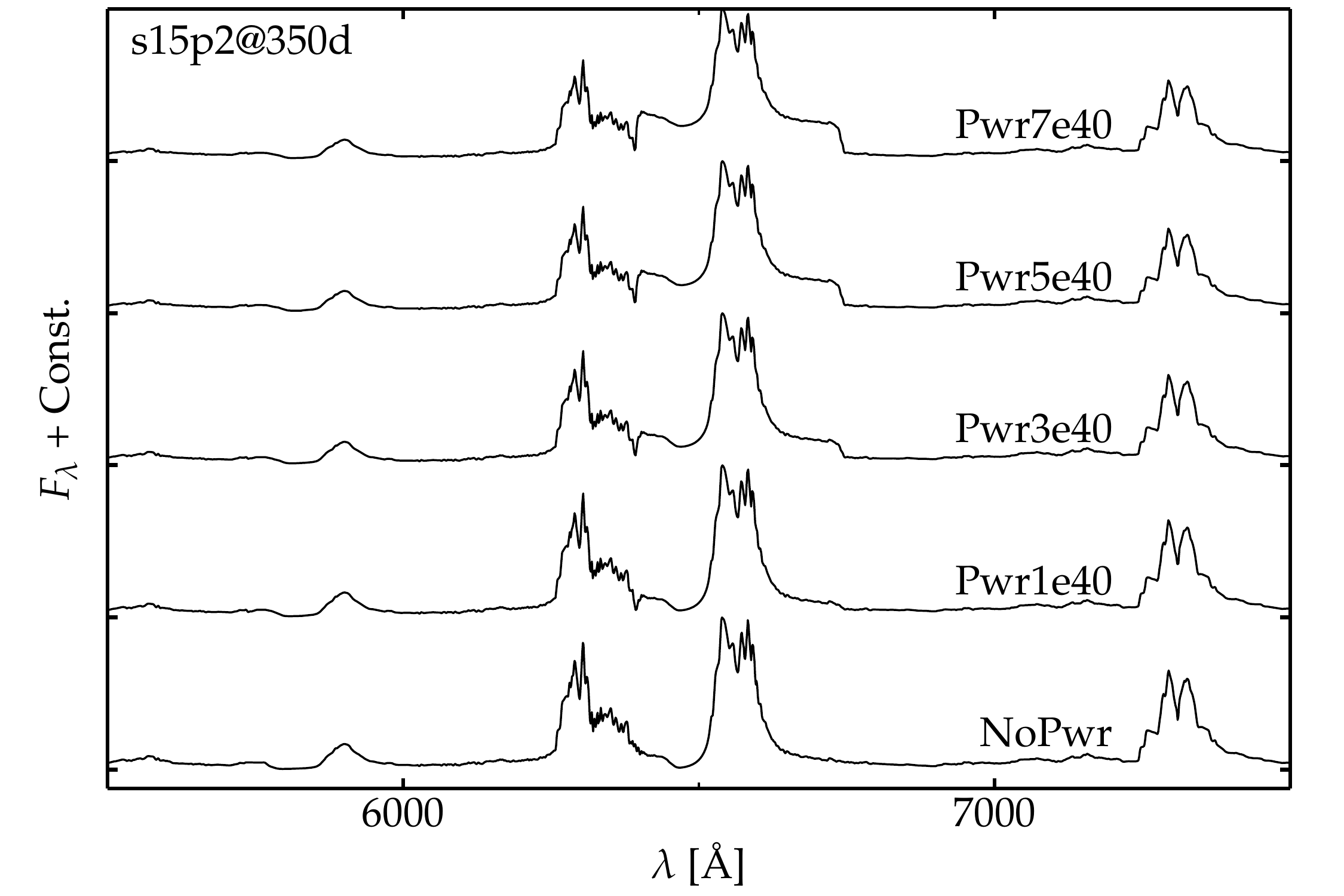}
\vspace{-0.4cm}
\caption{Spectral montage of the H$\alpha$ region for model s15p2 at 350\,d with shock power increasing upward from zero (label NoPwr; bottom) to $7 \times 10^{40}$\,\ergs\ (label Pwr7e40; top). Some of these models are used for comparison to observations in Section~\ref{sect_obs}. A broader range of shock powers covering up to $10^{43}$\,\ergs\ is used in \citet{DH_interaction_22}.
\label{fig_other_pwr}
}
\end{figure}

 \begin{figure}
\centering
\includegraphics[width=\hsize]{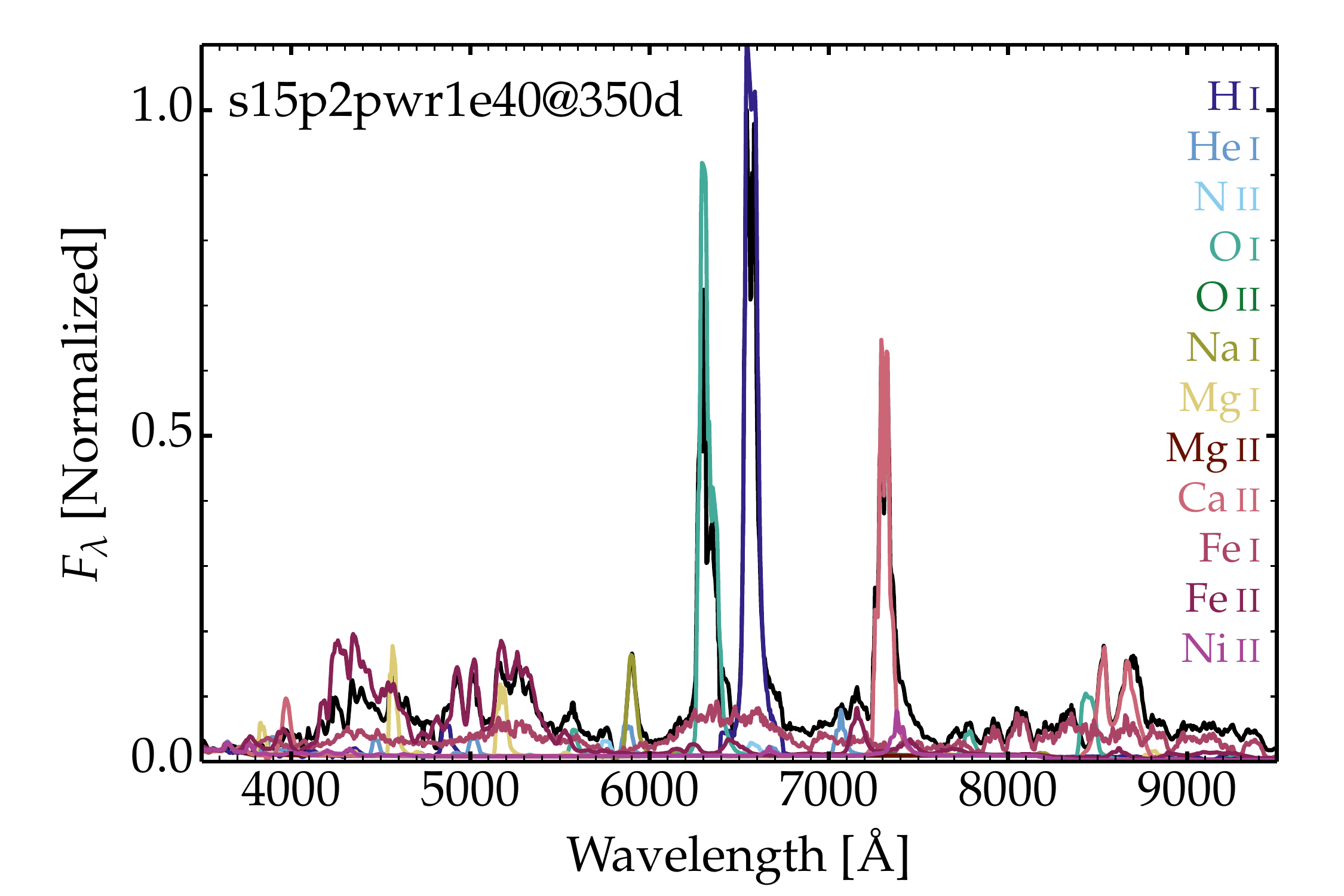}
\includegraphics[width=\hsize]{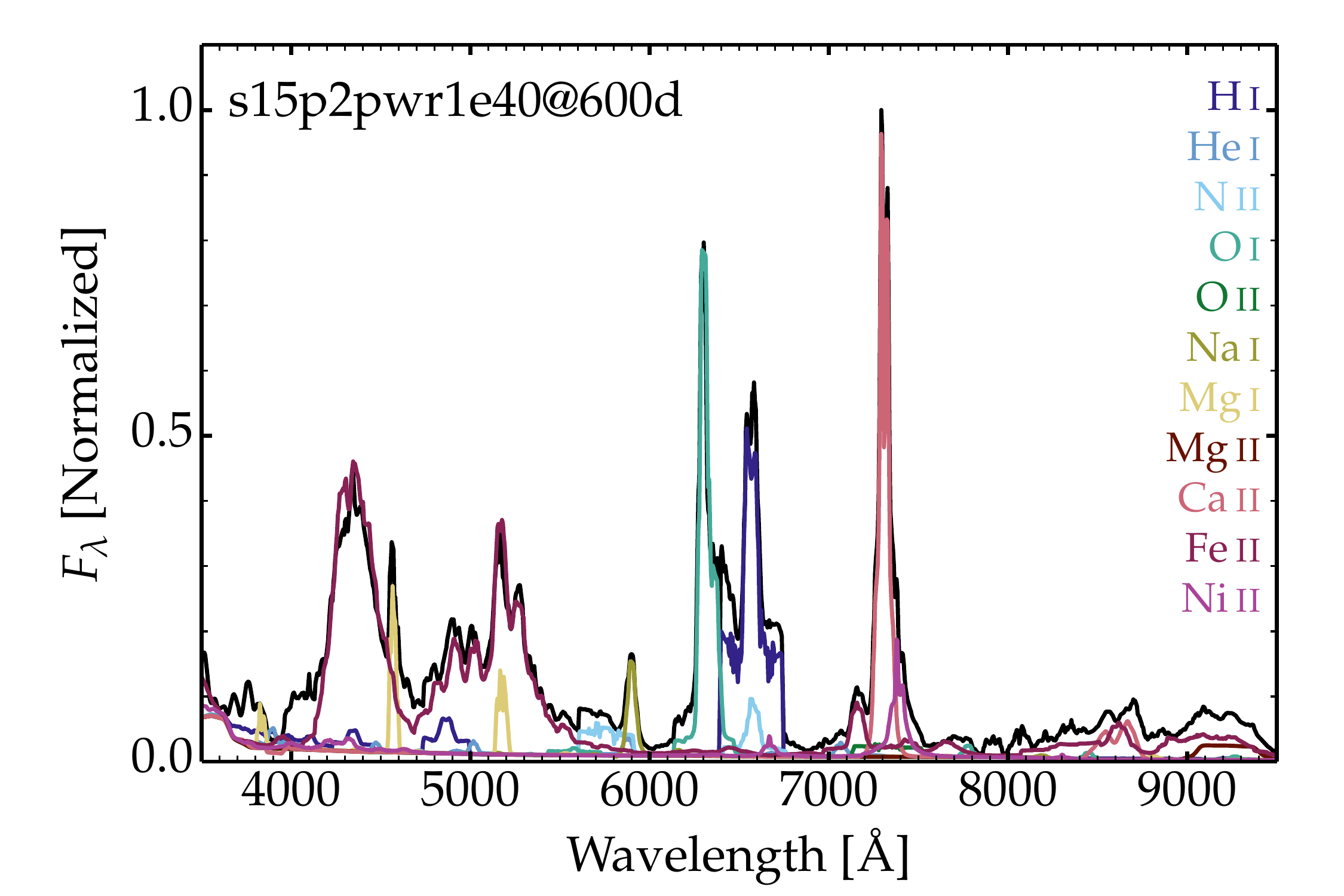}
\includegraphics[width=\hsize]{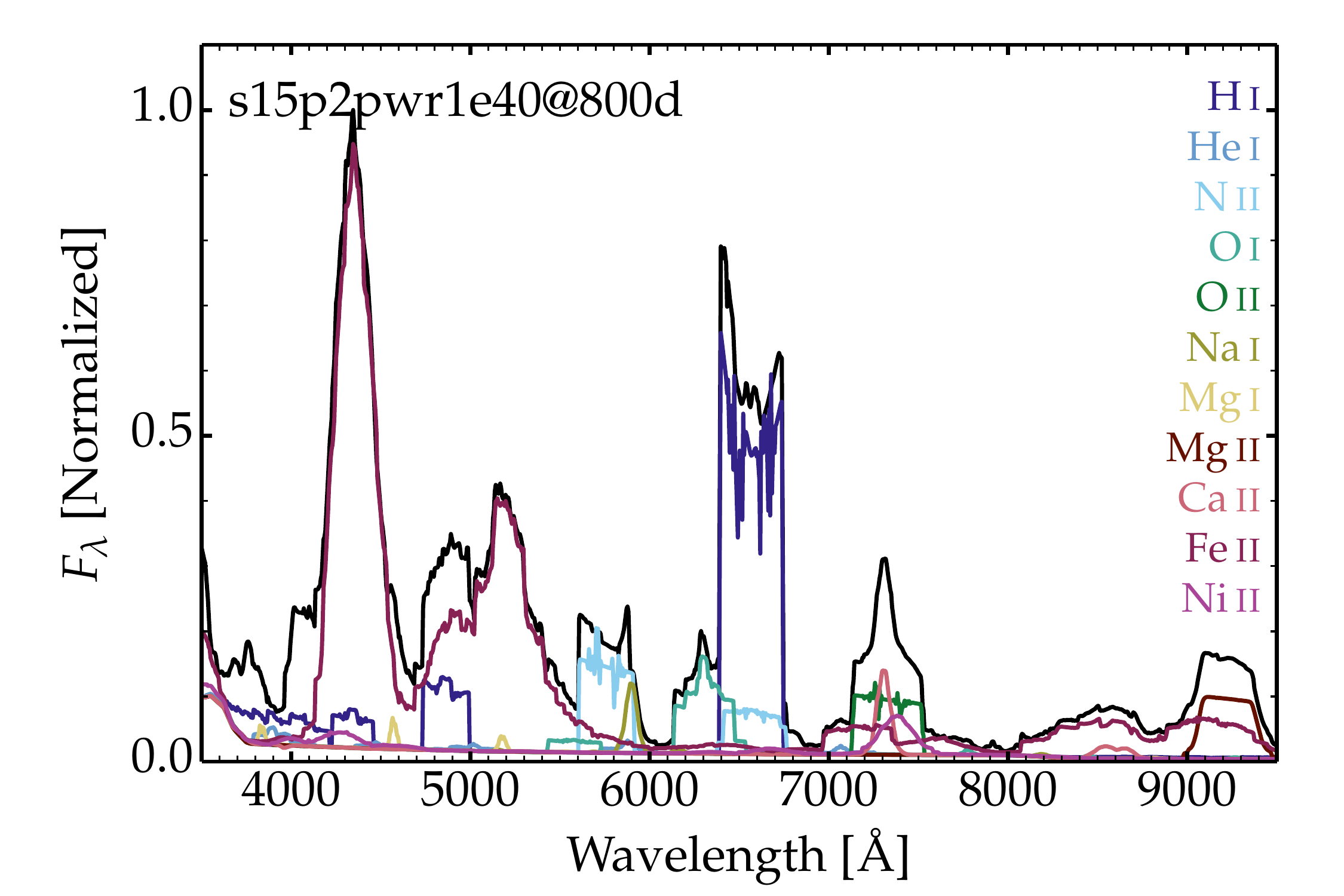}
\vspace{-0.4cm}
\caption{Line contributions to the full spectrum for model s15p2pwr1e40 at 350 (top), 600 (middle), and 800\,d (bottom). In each panel, we show the total normalized flux density as a black line together with the flux from individual ions (the same normalization is applied to all spectra) including H\one, He\one, N\two, O\one, O\two, Na\one, Mg\one, Mg\two, Ca\two, Fe\one, Fe\two, and Ni\two.
\label{fig_spec_ions}
}
\end{figure}

\end{document}